\documentclass[twocolumn,showpacs,preprintnumbers]{revtex4}
\usepackage{graphicx}% Include figure files
\usepackage{dcolumn}% Align table columns on decimal point
\usepackage{bm}% bold math
\usepackage{epsf}
\usepackage{subfigure}
\usepackage{epstopdf}
\usepackage{amsmath}
\usepackage{amssymb}

\newcommand{\bra}[1]{\langle #1|}
\newcommand{\ket}[1]{|#1\rangle}
\newcommand{\braket}[2]{\langle #1|#2\rangle}

\newcommand{\mean}[1]{\langle #1 \rangle}

\renewcommand{\i}{{\rm i}}
\newcommand{\e}{{\rm e}}
\begin{document}
\title{Continuous time random walk for open systems:\\ 
Fluctuation theorems and counting statistics}
\author{Massimiliano Esposito}
\altaffiliation[Also at \; ]{Center for Nonlinear Phenomena and Complex Systems,
Universite Libre de Bruxelles, Code Postal 231, Campus Plaine, B-1050 Brussels, Belgium.}
\author{Katja Lindenberg}
\affiliation{Department of Chemistry and Biochemistry and Institute for Nonlinear Science, University of California,
San Diego, La Jolla, CA 92093-0340, USA}

\date{\today}

\begin{abstract}
We consider continuous time random walks (CTRW) for open systems that exchange 
energy and matter with multiple reservoirs. 
Each waiting time distribution (WTD) for times between steps
is characterized by a positive parameter $\alpha$, which is set to 
$\alpha=1$ if it decays at least as fast as $t^{-2}$ at long times
and therefore has a finite first moment.  A WTD with $\alpha<1$ decays
as $t^{-\alpha-1}$.
A fluctuation theorem for the trajectory quantity $R$, defined as the logarithm 
of the ratio of the probability of a trajectory and the probability of 
the time reversed trajectory, holds for any CTRW. 
However, $R$ can be identified as a trajectory entropy change
only if the WTDs have $\alpha=1$ and satisfy separability (also called
``direction time independence").
For nonseparable WTDs with $\alpha=1$, $R$ can only be identified 
as a trajectory entropy change at long times, and a fluctuation theorem for the 
entropy change then only holds at long times.
For WTDs with $0<\alpha<1$ no meaningful fluctuation theorem can be derived.
We also show that the (experimentally accessible) $n$th moments of the energy and matter transfers
between the system and a given reservoir grow as $t^{n \alpha}$ at long times.
\end{abstract}

\maketitle
%%%%%%%%%%%%%%%%%%%%%%%%%%%%%%%%%%%%%%%%%%%%%%%%%%%%%%%%%%%%%%%%%%%%%%
\section{Introduction}\label{intro}

It has long been clearly understood that the statement of the Second Law of Thermodynamics
concerning the increase in entropy in an isolated system as it goes to equilibrium refers only to the
average behavior, but this was sufficient as long as one dealt only with macroscopic systems
characterized by
extremely narrow ensemble distributions with fluctuations that were essentially never observed.
More recently, with the ability to quantitatively monitor systems on the extremely small scales
of single molecules and quantum dots, it is possible to study fluctuations around the average
behavior. \emph{Fluctuation theorems} that hold arbitrarily far from equilibrium have thus
become subject to experimental verification~\cite{Bustamante02,
Bustamante04,Bustamante05,Evans05,Seifert05exp,Seifert06exp,Gaspard07exp}. 
These theorems in general deal with the ratio of the probabilities of a given system trajectory 
and that of its time reversed trajectory, either as the system goes to equilibrium or as it 
evolves to a steady state under the action of nonequilibrium constraints imposed on the system.  
From this one can calculate, for example, the relative probabilities that the entropy
of an isolated system away from thermodynamic equilibrium will spontaneously increase 
or decrease over a given period of time.  
The ratio is essentially infinite in a macroscopic system away from equilibrium and is unity
due to fluctuations in equilibrium, but in sufficiently small systems away from equilibrium 
it is merely large (and experimentally accessible) rather than infinite.  

Fluctuation theorems can take different forms depending on the specific 
problem under consideration, but they are all ultimately connected to the 
probabilistic asymmetry of system trajectories and time reversed trajectories.
Equilibrium corresponds to the situation where the symmetry 
is restored as stated by the principle of microreversibility. 
Fluctuation theorems have been formulated for a wide range of dynamics such 
as driven Hamiltonian dynamics~\cite{Jarzynski97,VandenBroeck07,CleurenVDBroeckKawai}, stochastic
dynamics~\cite{Kurchan98,Lebowitz99,Crooks,Seifert05,AndrieuxGaspard07a,EspositoHarbola07PRE},
deterministic thermostated dynamics~\cite{Evans,Gallavotti}, and even quantum 
dynamics~\cite{Kurchan00,HTasaki00,Mukamel03,TalknerHanggi07,EspositoHarbola07,CleurenVDBroeck}.
Here we focus on stochastic dynamics, in an effort to explore the validity of fluctuation 
theorems beyond the stochastic dynamics that have been considered to date. 

In this narrower context of stochastic dynamics, most previous studies of fluctuation theorems 
have focused on systems described by Markovian master equations or Fokker-Plank equations.  
Recently there have been some efforts to investigate fluctuation theorems for systems 
described by nonlinear generalized Langevin equations~\cite{Kurchan05,Ohta07} 
with an external driving force as a nonequilibrium constraint.  
Our focus is on nonequilibrium systems described by continuous time random 
walks (CTRW)~\cite{Montroll65,Lax73,Kehr87,Georges90,Lindenberg71,Hughes} in which 
transitions between microscopic states may be caused by more than one mechanism.
The nonequilibirum constraint is imposed when these mechanisms have different statistical properties
such as, for example, through contact with two heat baths maintained at different temperatures.  
In general, identifying such nonequilibrium constraints may itself be 
complicated~\cite{AndrieuxGaspard07a,Schnakenberg}, and we will here 
explicitly point to these differences.

We pose the following question: What properties of a CTRW are necessary 
for an entropy fluctuation theorem to be valid under such nonequilibrium constraints?  
We note, for example, that CTRWs are known to display aging \cite{Barkai03,Allegrini03,Sokolov} 
as well as nonergodic phenomena \cite{Barkai06,Barkai07} which may significantly alter the 
behavior of the system under time reversal and prevent a fluctuation theorem from being
satisfied.  At the same time, CTRWs under certain conditions reduce to Markovian 
master equations which are known to satisfy fluctuation theorems.
CTRWs therefore provide a good framework to study the validity of fluctuation theorems.
In particular, our results will hopefully contribute clarification to recent observations 
of anomalous statistics in the nonequilibrium fluctuations of single molecules and quantum
dots~\cite{Silbey02,Bouchaud03,Yang03,KlafterExp05}.

A second purpose of this paper is the formulation of a general framework for the
calculation of (experimentally accessible) counting statistics of events associated with
a given mechanism. Examples of such events might involve particle or energy transfer.
To accomplish this we use a method based on the propagation of the generating function 
associated with the probability distribution of the events, in the spirit of the method 
used for Markovian master equations~\cite{Lebowitz99,EspositoHarbola07PRE}.
This will allow us to investigate the long-time behavior of the 
moments of the distribution associated with the counting statistics.  

Our basic CTRW model is constructed as follows.
We consider a stochastic dynamics between microscopic states $m$ of a system
with a finite number of states.  The transitions between states
may be due to different mechanisms $\nu$.  For example, we will subsequently consider a system in
which each microscopic state $m$ is characterized by a number $N_m$ of particles and an energy 
$\epsilon_m$, and where the transitions between the $m$'s are triggered by different reservoirs
(heat baths) $\nu$. Suppose that the system arrives at state $m'$ at a given time and that its next
jump is to state $m$ at a time $t$ later via mechanism $\nu$.  The distribution of waiting times
(WTD) for this to occur is denoted by $\psi_{mm'}^{(\nu)}(t)$, with other related
quantities specified in more detail subsequently. 
We focus on waiting time distributions whose long-time behavior is reflected in the small-$s$
Laplace transform
\begin{equation}
\tilde{\psi}_{mm'}^{(\nu)}(s) \underset{s \to 0}{=}
P_{mm'}^{(\nu)}- B_{mm'}^{(\nu)} s^{\alpha} ,
\label{s0WTDpsigeneral}
\end{equation}
where $\tilde{f}(s) \equiv \int_0^\infty dt \e^{-st} f(t)$ and  $0 < \alpha \leq 1$.
The $B_{mm'}^{(\nu)}$ are elements of an arbitrary matrix.
A detailed discussion surrounding this choice can be found in~\cite{Shlesinger}.
When $0<\alpha<1$ the long-time decay of the WTDs is then of the 
power law form $\psi_{mm'}^{(\nu)}(t) \sim t^{-\alpha-1}$. When 
$\alpha=1$ the decay is at least as fast as $1/t^2$ but may be faster.

%%%%%%%%%%%%%%%%%%%%%%%%%%%%%%%%%%%%%%%%%%%%%%

In Sec.~\ref{CTRW} we present the CTRW model for an open system driven by different
mechanisms described by different statistical properties, and formally express the 
probability that the system is in state $m$ at time $t$.  
In Sec.~\ref{GME} we derive the generalized master equation 
satisfied by this probability and study its long time behavior.   
In Sec.~\ref{GFcurrentsec} we present a generating function formalism 
to calculate the probability distribution of heat and matter transfers
which is used to study the long time behavior of the moments of the distribution. 
The principal results of this work, namely, the conditions for the validity of fluctuation
theorems, are presented in Sec.~\ref{Ptraj}.
In particular, we show that fluctuation theorems for the entropy change can only be obtained if
$\alpha=1$, that is, if the WTDs decay at least as fast as $t^{-2}$ at long times. 
Furthermore, even in this case the entropy change can be expressed as a familiar ratio
of the probability of a trajectory and its time reversed trajectory only if the WTDs satisfy 
constraints of separability.
%%%%%%%%%%%%%%%%%%%%%
%%%%%%%%%%%%%%%%%%%%%
%In part~\ref{SSFlucTheo} we use the generating function formalism together with
%large deviation theory to derive a steady-state fluctuation theorem for the 
%entropy change which is valid for WTDs with $\alpha=1$.
%In part~\ref{TRev} a fluctuation theorem for the quantity R, defined as 
%the logarithm of the ratio of the probability of a trajectory and its time 
%reversed trajectory, and show that R can be interpreted as a trajectory 
%entropy change only for separable WTDs with $\alpha=1$.  
%The connection between these two type of fluctuation theorems is also discussed.
A summary of results and some concluding remarks are presented in Sec.~\ref{conc}.    

%%%%%%%%%%%%%%%%%%%%%%%%%%%%%%%%%%%%%%%%%%%%%%%%%%%%%%%%%%%%%%%%%%%%%%
\section{Continuous time random walks for open systems}\label{CTRW}

Our goal in this section is to construct the probability that the system will be found in a
particular state at time $t$.  Suppose that the system arrives at state $m'$ at time zero,
and that its next jump is to state $m$ at time $t$ via mechanism $\nu$.  The
waiting time distribution (WTD) $\psi_{mm'}^{(\nu)}(t)$ for this event introduced earlier
satisfies the normalization condition
\begin{equation}
\int_{0}^{\infty} d\tau \sum_{m,\nu} \psi_{mm'}^{(\nu)}(\tau) = 1.
\label{normWTD}
\end{equation}
For convenience, we define $\psi_{mm}^{(\nu)}(\tau) \equiv 0$.
The probability that no transition occurs up to time $t$ after arrival at
$m'$ at time zero is
\begin{eqnarray}
\phi_{m'}(t) &=& \sum_{m,\nu} \int_{t}^{\infty} d\tau \; \psi_{mm'}^{(\nu)}(\tau) \nonumber\\
&=&1-\sum_{m,\nu} \int_{0}^{t} d\tau \; \psi_{mm'}^{(\nu)}(\tau) \;.
\label{nojumpWT}
\end{eqnarray}
We see that by construction $\phi_{m'}(\infty)=0$, that is, a jump eventually occurs with certainty.
We define the auxiliary distributions
\begin{equation}
\psi_{m'}^{(\nu)}(t) \equiv \sum_{m} \psi_{mm'}^{(\nu)}(t), \quad
\psi_{m'}(t) \equiv \sum_{\nu} \psi_{m'}^{(\nu)}(t) \;. \label{WTDwithoutarrival}
\end{equation}
$\psi_{m'}^{(\nu)}(t)$
%[$\psi_{m'}(t)$]
is the waiting time distribution of
the first jump from state $m'$ to any other state via a mechanism $\nu$,
and $\psi_{m'}(t)$ is the waiting time distribution of the first jump from state $m'$ to any other
state regardless of the mechanism, given that arrival at $m'$ occurred at time zero.
We also define
\begin{equation}
P_{mm'}^{(\nu)} \equiv \int_{0}^{\infty} d\tau \psi_{mm'}^{(\nu)}(\tau), \quad
P_{mm'} \equiv \sum_{\nu} P_{mm'}^{(\nu)} .
\label{defT}
\end{equation}
Here $P_{mm'}^{(\nu)}$
is the probability that, being at $m'$, the
next jump will be from state $m'$ to $m$ via mechanism $\nu$, and $P_{mm'}$ 
is the probability that the jump will be from $m'$ to $m$ irrespective of the mechanism. 
Note that by definition $P_{mm} \equiv 0$.
We finally define the probability $f_{m'}^{\nu}$ that the next jump from $m'$ will be due to
mechanism $\nu$,
\begin{equation}
f_{m'}^{\nu} \equiv \sum_{m} P_{mm'}^{(\nu)}.
\label{deff}
\end{equation}
The normalization condition (\ref{normWTD}) implies 
\begin{equation}
\int_{0}^{\infty} d\tau \psi_{m}(t)=1, \quad
\sum_m P_{mm'}=1, \quad
\sum_{\nu} f_{m'}^{(\nu)} = 1 .
\end{equation}

%In~\cite{QianWang} it is argued that a necessary condition for a CTRW to satisfy 
%microreversibility is the separability of the waiting time distribution into the 
%product of a waiting time portion that depends only on the originating state, 
%and a transition matrix that connects a given initial and final state:
%\begin{equation}
%\psi_{mm'}^{(\nu)}(t) = P_{mm'}^{(\nu)} \; \psi_{m'}(t).
%\label{DTI}
%\end{equation}
%We note that it is implicitly assumed that the WTDs have a finite first moment (see below).
%Another necessary condition is that the system also obeys detailed balance (which 
%it cannot do if different transition mechanisms have different statistics, see below).
%Together, these two necessary condition constitute a sufficient condition.
%(The separability condition is called ``direction time independence" 
%by Qian and Wang~\cite{QianWang,WangQian}.) 
%We will see that separability also plays a role in the derivation of a fluctuation theorem. 

The determining feature of the problem for our purposes is the behavior of the first 
moment of the waiting time distribution $\psi_{mm'}^{(\nu)}(t)$ and, in particular, 
whether it is finite or infinite.  
This first moment is just the average time that the system remains in state 
$m'$ before jumping to $m$ via mechanism $\nu$:  
\begin{equation}
t_{mm'}^{(\nu)} \equiv \int_{0}^{\infty} d\tau \; \tau \; \psi_{mm'}^{(\nu)}(\tau).
%= -\frac{d}{ds} \tilde{\psi}_{mm'}^{(\nu)}(s=0) \;.
\label{tmoyenEntre2m}
\end{equation}
Associated mean waiting times are given by
\begin{equation}
t_{m'}^{(\nu)} \equiv \sum_{m} t_{mm'}^{(\nu)}, \quad
t_{m'} \equiv \sum_{\nu} t_{m'}^{(\nu)} .
\end{equation}
The first is the average time that the system remains in state $m'$ without 
jumping anywhere else via mechanism $\nu$.  
The second is the average time that the system remains at $m'$ without
making any jumps at all by any mechanism.  
%We focus on waiting time distributions whose long-time
%behavior is reflected in the small-$s$ Laplace transform
%\begin{equation}
%\tilde{\psi}_{mm'}^{(\nu)}(s) \underset{s \to 0}{=} 
%P_{mm'}^{(\nu)}- B_{mm'}^{(\nu)} s^{\alpha} ,
%\label{s0WTDpsigeneral}
%\end{equation}
%where $\tilde{f}(s) \equiv \int_0^\infty dt \e^{-st} f(t)$ and  $0 < \alpha \leq 1$.  
%The $B_{mm'}^{(\nu)}$ are elements of an arbitrary matrix.   
%A detailed discussion surrounding this choice can be found in~\cite{Shlesinger}. 
%Since the Laplace transform of~(\ref{nojumpWT}) reads
%\begin{equation}
%\tilde{\phi}_{m}(s) = \frac{1}{s} \sum_{m'} \big( \delta_{mm'}-\tilde{\psi}_{m'm}(s) \big) 
%= \frac{1}{s} \big( 1-\tilde{\psi}_{m}(s) \big),
%\label{laplacephi}
%\end{equation}
%it follows that
%\begin{equation}
%\tilde{\phi}_{m}(s) \underset{s \to 0}{=} B_{m} s^{\alpha-1},
%\label{s0WTDphigeneral}
%\end{equation}
%where $B_{m}=\sum_{m',\nu} B_{m'm}^{(\nu)}$.
%The important feature in these long time--small $s$ behaviors that will distinguish 
%between the finite and infinite mean waiting times, is the value of the exponent $\alpha$. 
If $\alpha < 1$, all moments of the waiting time distribution (including the first
moments or mean waiting times) are divergent, whereas for 
$\alpha = 1$ the first moments $t_{mm'}^{(\nu)}$ 
are finite.  In this case, from~(\ref{s0WTDpsigeneral}) and~(\ref{tmoyenEntre2m}) 
it follows that $B_{mm'}^{(\nu)}=t_{mm'}^{(\nu)}$. 

Suppose that we begin our observations at time $t=0$, at which time we find the system
in state $m'$.  Since in general we may not know when the jump occurred that brought the system
to that state, we need to distinguish the waiting time distribution of the first jump after time zero
from that of subsequent jumps. We mark this first waiting time distribution with a prime,
$\psi_{mm'}'^{(\nu)}(t)$. The probability that no transition away from $m'$ occurs up to time $t$
must be similarly distinguished, $\phi_m'(t)$.  The primed functions are equal to the unprimed ones
only if a jump occurred exactly at time zero or if the WTDs decay exponentially. If there is no
information about when the last jump
before time $t=0$ occurred, then the primed functions can be related to the unprimed ones
\emph{only} if the mean waiting times $t_{mm'}^{(\nu)}$ are finite.  
In this case, an average over the uncertain past yields 
\begin{eqnarray}
\psi_{mm'}'^{(\nu)}(t) &=&
\frac{\int_{-\infty}^{0} d\tau \; \psi_{mm'}^{(\nu)}(t-\tau)}
{\sum_{m,\nu} \int_{0}^{\infty} dt \int_{-\infty}^{0} d\tau\; \psi_{mm'}^{(\nu)}(t-\tau)} \nonumber \\
&=& \frac{\int_{-\infty}^{0} d\tau \;  \psi_{mm'}^{(\nu)}(t-\tau)}{t_{m'}}
= \frac{\int_{t}^{\infty} d\tau \;  \psi_{mm'}^{(\nu)}(\tau)}{t_{m'}} .\nonumber\\
&&
\label{WTfirstjump}
\end{eqnarray}
%The denominator ensures that (\ref{normWTD}) is satisfied. 
Note that summing Eq.~(\ref{WTfirstjump}) over $m$ and $\nu$ and using Eq.~(\ref{nojumpWT}) leads to
\begin{equation}
\psi_{m'}'(t)=\frac{\phi_{m'}(t)}{t_{m'}} .
\label{WTfirstjumpbis}
\end{equation}
Similarly, 
\begin{equation}
\phi_{m'}'(t)
%&=& \frac{\int_{0}^{\infty} d\tau \phi_{m'}(t+\tau)}{\int_{0}^{\infty} d\tau \phi_{m'}(\tau)}
= \frac{\int_{t}^{\infty} d\tau \phi_{m'}(\tau)}{t_{m'}} .
%\nonumber\\ 
%&=& \sum_{m,\nu} \int_{t}^{\infty} d\tau \psi_{mm'}'^{(\nu)}(\tau) \;,
\label{WTnojumpfirstjump}
\end{equation}
%where we used (\ref{WTfirstjumpbis}) to get the last line.

%In Laplace space, integrating by part and using the initial value theorem 
%($f(0+)=\lim_{s \to \infty} s \tilde{f}(s)$), (\ref{WTfirstjump}) and (\ref{WTnojumpfirstjump}) become
%\begin{eqnarray}
%&&\tilde{\psi}_{mm'}'^{(\nu)}(s)=\frac{1}{s \; \bar{t}_{m'}} 
%\big(P_{mm'}^{(\nu)}-\tilde{\psi}_{mm'}^{(\nu)}(s) \big) \nonumber \\
%&&\tilde{\phi}_{m}'(s)=\frac{1}{s \; \bar{t}_{m}} \big(\bar{t}_{m}-\tilde{\phi}_{m}(s) \big) \;.
%\label{WTnojumpfirstjumpLaplace}
%\end{eqnarray}
%Using (\ref{s0WTDpsigeneral}) and (\ref{s0WTDphigeneral}) with $\alpha=1$ 
%and $B_{mm'}=\bar{t}_{mm'}$ and $B_{m}=\bar{t}_{m}$, the long time 
%behavior of (\ref{WTnojumpfirstjumpLaplace}) is given by
%\begin{eqnarray}
%\tilde{\psi}_{mm'}'^{(\nu)}(s) \stackrel{s \to 0}{=} 
%\frac{\bar{t}_{mm'}^{(\nu)}}{\bar{t}_{m'}}+{\cal O}(s) \;, \label{asympphi2}
%\end{eqnarray}
%and $\lim_{s \to 0} \tilde{\phi}_{m}'(s)$ is zero [constant] 
%if the second moment of the WTD is divergent [finite]. 
%Using (\ref{DTI}), (\ref{tmoyenEntre2m}) and (\ref{tmoyen}), we find that
%for DTI WTD $\bar{t}_{mm'}^{(\nu)}= P_{mm'}^{(\nu)} \bar{t}_{m'}$, so that in
%the long time limit of (\ref{asympphi2}) and (\ref{s0WTDpsigeneral}) become identical.\\
%It is worth noting that for DTI exponential WTD, this averaging procedure is equivalent to 
%assume that a jump occurred at time zero. Indeed, if $\psi_{m}(t)=\e^{-t/\bar{t}_m}/\bar{t}_m$, 
%$\phi_{m}(t)=\e^{-t/\bar{t}_m}$, so that $\psi_{m}'(t)=\psi_{m}(t)$.\\

It is straightforward to construct an integral equation for $\rho_m(t)$, 
the probability that the system is in state $m$ at time $t$.  
For this purpose we also define $\eta_m(t)$, the probability
that the system jumps onto state $m$ at time $t$.  
The following integral CTRW relations are evident~\cite{Montroll65,Lax73,Kehr87,Georges90,Hughes}:
\begin{eqnarray}
\eta_m(t) &=& \sum_{m',\nu} \psi_{mm'}'^{(\nu)}(t)  \rho_{m'}(0) \nonumber \\
&&+ \sum_{m',\nu} \int_{0}^{t} d\tau \psi_{mm'}^{(\nu)} (t-\tau) \eta_{m'}(\tau) ,
\label{CTRW1}
\end{eqnarray}
and 
\begin{equation}
\rho_m(t) = \phi_{m}'(t) \rho_{m}(0) 
+ \int_{0}^{t} d\tau \phi_{m}(t-\tau) \eta_{m}(\tau) .
\label{CTRW2}
\end{equation}
Since these are convolutions, it is easiest to solve for the Laplace transform
$\tilde{\rho}_m(s)$ of the desired
probability $\rho_m(t)$.  The Laplace transforms of these two relations are
\begin{equation}
\tilde{\eta}_m(s) = \sum_{m',\nu} \bigg( \tilde{\psi}_{m,m'}'^{(\nu)}(s)  \rho_{m'}(0) 
+ \tilde{\psi}_{m,m'}^{(\nu)} (s) \tilde{\eta}_{m'}(s) \bigg) ,
\label{CTRW1lap}
\end{equation}
and 
\begin{equation}
\tilde{\rho}_m(s) = \tilde{\phi}'_{m}(s) \rho_{m}(0) + \tilde{\phi}_{m}(s) \tilde{\eta}_{m}(s) .
\label{CTRW2lap}
\end{equation}
The solution is most neatly expressed in terms of the matrices $\Phi$ and $\Psi$ with matrix elements
\begin{equation}
[\tilde{\Psi}(s)]_{mm'} \equiv \tilde{\psi}_{m,m'}'(s), \quad
[\tilde{\Phi}(s)]_{mm'} \equiv \delta_{mm'} \tilde{\phi}_{m}(s),
\label{CTRWformalsol1}
\end{equation}
and the vectors $\ket{\tilde{\eta}(s)}$ and $\ket{\tilde{\rho}(s)}$ with elements
$\tilde{\eta}_m(s)$ and $\tilde{\rho}_m(s)$.  The solution of Eq.~(\ref{CTRW1lap}) is
\begin{equation}
\ket{\tilde{\eta}(s)} = \big({I}-{\tilde{\Psi}}(s) \big)^{-1} {\tilde{\Psi}}'(s) \ket{\rho(0)} ,
\label{CTRWformalsol2}
\end{equation}
where $I$ is the identity matrix.
Using~(\ref{CTRW2lap}), the formal solution of the CTRW then is
\begin{eqnarray}
\ket{\tilde{\rho}(s)} = \bigg({\tilde{\Phi}}'(s) + {\tilde{\Phi}}(s)  
\big({I}-{\tilde{\Psi}}(s) \big)^{-1} {\tilde{\Psi}}'(s) \bigg) \ket{\rho(0)} .
\label{CTRWformalsol3}
\end{eqnarray}

\iffalse
If $\psi_{m,m'}^{(\nu)}(t)=P_{mm'}^{(\nu)} \psi^{(\nu)}(t)$, we get the simplified form
\begin{eqnarray}
\label{CTRWformalsol3Simplif}
\ket{\tilde{\rho}(s)} &=& \big(\hat{I}- \tilde{\psi}(s) \hat{P} \big)^{-1} \\
&&\bigg(1-\tilde{\psi}(s)+\big(\tilde{\psi}'(s)
-\tilde{\psi}(s) \big) \hat{P} \bigg) \ket{\rho(0)} \;.\nonumber
\end{eqnarray}
\fi

%%%%%%%%%%%%%%%%%%%%%%%%%%%%%%%%%%%%%%%%%%%%%%%%%%%%%%%%%%%%%%%%%%%%%%
\section{Generalized master equation}\label{GME}

In this section we construct the generalized master equation for the probability that the 
system is in state $m$ at time $t$.  This equation facilitates the exploration of the behavior
of this probability at long times. 
Using the Laplace transform of~(\ref{nojumpWT}),
\begin{equation}
\tilde{\phi}_{m}(s) = \frac{1}{s} \sum_{m'} \big( \delta_{mm'}-\tilde{\psi}_{m'm}(s) \big)
= \frac{1}{s} \big( 1-\tilde{\psi}_{m}(s) \big),
\label{laplacephi}
\end{equation}
we can rewrite (\ref{CTRW2lap}) as
\begin{eqnarray}
\tilde{\rho}_m(s) = \tilde{\phi}'(s) \rho_{m}(0) + \frac{1}{s} \tilde{\eta}_{m}(s) 
- \frac{1}{s} \sum_{m',\nu} \tilde{\psi}_{m'm}^{(\nu)}(s) \tilde{\eta}_{m}(s) \;.
\nonumber\\ \label{CTRW3lap}
\end{eqnarray}
Inserting (\ref{CTRW1lap}) in the second term of (\ref{CTRW3lap})
%, we find
%\begin{eqnarray}
%\tilde{\rho}_m(s) &=& \frac{1}{s} \rho_{m}(0) + \frac{1}{s} \sum_{m',\nu} \\
%&&\hspace{-1.6cm} \bigg( \tilde{\psi}_{mm'}^{(\nu)}(s) \big( \tilde{\eta}_{m'}(s) + \rho_{m'}(0) \big) 
%- \tilde{\psi}_{m'm}^{(\nu)}(s) \big( \tilde{\eta}_{m}(s) + \rho_{m}(0) \big) \bigg) \nonumber
%\label{CTRW4lap}
%\end{eqnarray}
and using Eq.~(\ref{CTRW2lap}), we obtain
\begin{eqnarray}
&&s \tilde{\rho}_m(s) - \rho_{m}(0) = \tilde{\mathcal{I}}(s)  \nonumber \\
&&\hspace{1.1cm}+ \sum_{m',\nu} \bigg( \tilde{W}_{mm'}^{(\nu)}(s) \tilde{\rho}_{m'}(s) 
- \tilde{W}_{m'm}^{(\nu)}(s) \tilde{\rho}_{m}(s) \bigg), \nonumber \\
&&
\label{GeneralizedMELaplace}
\end{eqnarray}
where the transition matrix elements are given by
\begin{equation}
\tilde{W}_{mm'}^{(\nu)}(s) = \frac{\tilde{\psi}_{mm'}^{(\nu)}(s)}{\tilde{\phi}_{m'}(s)}
\label{rate}
\end{equation}
and 
\begin{eqnarray}
\tilde{\mathcal{I}}(s) &\equiv& \sum_{m',\nu} \bigg( \tilde{W}_{mm'}^{(\nu)}(s) 
\big( \tilde{\phi}_{m'}(s) - \tilde{\phi}_{m'}'(s) \big) \rho_{m'}(0) \nonumber \\
&&\hspace{0.5cm}-  \tilde{W}_{m'm}^{(\nu)}(s) 
\big( \tilde{\phi}_{m}(s) - \tilde{\phi}_{m}'(s)\big) \rho_{m}(0) \bigg).
\label{inhomog}
\end{eqnarray}
The inhomogeneous term $\tilde{\mathcal{I}}(s)$ in Eq.~(\ref{GeneralizedMELaplace}) 
thus depends on the initial condition.
It vanishes if a jump occurs at time zero or if the waiting time distributions are 
exponential because then $\phi_{m}'(t)=\phi_{m}(t)$.
Upon inverse Laplace transformation we arrive at the generalized master equation 
\begin{eqnarray}
&&\dot{\rho}_m(t) = \mathcal{I}(t) + \sum_{m',\nu} \int_{0}^{t} d\tau 
\label{GeneralizedME}\nonumber \\
&&\hspace{0.8cm} \times \bigg( W_{mm'}^{(\nu)}(\tau) \rho_{m'}(t-\tau) 
- W_{m'm}^{(\nu)}(\tau) \rho_{m}(t-\tau) \bigg). \nonumber\\ 
&&
\end{eqnarray}
The generalized master equation is clearly non-Markovian 
unless $\tilde{W}_{mm'}^{(\nu)}(s)$ is independent of $s$.  
This occurs, for example, for separable distributions
$\psi_{mm'}^{(\nu)}(t) = P_{mm'}^{(\nu)} \; \psi_{m'}(t)$
with $\psi_{m}(t)=\e^{-t/t_m}/t_m$. 
Indeed, in this case $\tilde{W}_{mm'}^{(\nu)}(s)=P_{mm'}^{(\nu)}/t_{m'}$.

Of interest for our purposes is the long-time behavior of the probability $\rho_m(t)$.  
Here we distinguish the case $\alpha=1$, associated with finite mean waiting times, 
from the case $0<\alpha<1$, associated with divergent mean waiting times.  

%%%%%%%%%%%%%%%%%%%%%%%%%%%%%
%%%%%%%%%%%%%%%%%%%%%%%%%%%%%
%%%%%%%%%%%%%%%%%%%%%%%%%%%%%
\subsection{Long time behavior for $\alpha=1$} \label{longtimeGME}

Consider first the long time behavior of the generalized 
master equation in the case $\alpha=1$.  From Eq.~(\ref{laplacephi}) it follows that
\begin{equation}
\tilde{\phi}_{m}(s) \underset{s \to 0}{=} B_{m} s^{\alpha-1},
\label{s0WTDphigeneral}
\end{equation}
where $B_{m}=\sum_{m',\nu} B_{m'm}^{(\nu)}$.
Since it follows from this result and from Eq.~(\ref{s0WTDpsigeneral}) 
that $B_{mm'}={t}_{mm'}$ and $B_{m}={t}_{m}$, Eq.~(\ref{rate}) then immediately leads to
\begin{equation}
\tilde{W}_{mm'}^{(\nu)}(0) = \lim_{s \to 0} \tilde{W}_{mm'}^{(\nu)}(s) 
= \frac{\tilde{\psi}_{mm'}^{(\nu)}(0)}{\tilde{\phi}_{m'}(0)}
= \frac{P_{mm'}^{(\nu)}}{{t}_{m'}}.
\label{SumrateMarkov}
\end{equation}
Since $\lim_{s \to 0} (\tilde{\phi}_{m'}(s) - \tilde{\phi}_{m'}'(s))$ 
is constant when $\alpha=1$, we find that $\lim_{s \to 0} \mathcal{I}(s)$ is also
constant [see Eq.~(\ref{inhomog})]. 
Therefore, using the final value theorem $f(\infty)=\lim_{s \to 0} s \tilde{f}(s)$ 
\cite{Hughes,Feller}, we find that
\begin{equation}
\lim_{t \to \infty} \mathcal{I}(t)=0 , \quad
\lim_{t \to \infty} W_{mm'}^{(\nu)}(t)=0 .
\label{IsecasseTlong}
\end{equation}
This means that at long times, the generalized master 
equation behaves like the Markovian master equation 
\begin{equation}
\label{MEfake}
\dot{\rho}_m(t) = \sum_{m',\nu}
\bigg( \tilde{W}_{mm'}^{(\nu)}(0) \rho_{m'}(t) 
- \tilde{W}_{m'm}^{(\nu)}(0) \rho_{m}(t) \bigg) .
\end{equation}
Defining the rate matrix $\boldsymbol{V} \equiv \sum_{\nu} \boldsymbol{V}^{(\nu)}$ where
\begin{eqnarray}
&&V_{mm'}^{(\nu)} \equiv \frac{P_{mm'}^{(\nu)}}{B_{m'}} \ \ {\rm for} \ \ m \neq m' \nonumber \\
&&V_{mm}^{(\nu)} \equiv - \sum_{m'(\neq m)} V_{m'm}^{(\nu)} , \label{ratebis}
\end{eqnarray}
we can rewrite (\ref{MEfake}) as $\ket{\dot{\rho}(t)} = \boldsymbol{V} \ket{\rho(t)}$.
Using the Perron-Frobenious theorem, all eigenvalues of $\boldsymbol{V}$
are negative aside from one which is zero.
The probability $\rho_m(t)$ therefore decays exponentially to a steady 
state solution $\rho_{m}^{\rm ss}$ that obeys the condition
\begin{equation}
\sum_{m',\nu} \big( \tilde{W}_{mm'}^{(\nu)}(0) \rho_{m'}^{\rm ss} 
- \tilde{W}_{m'm}^{(\nu)}(0) \rho_{m}^{\rm ss} \big) = 0.
\label{NESS}
\end{equation}
The steady state solution corresponds to equilibrium 
if the detailed balance condition
\begin{equation}
\tilde{W}_{mm'}^{(\nu)}(0) \rho_{m'}^{\rm eq} 
= \tilde{W}_{m'm}^{(\nu)}(0) \rho_{m}^{\rm eq} ,
\label{detailbalance}
\end{equation}
is satisfied, that is, if all fluxes between pairs of states associated 
with the different mechanisms $\nu$ become zero at equilibrium.
This would not be possible if the long-time statistics of the different mechanisms
were different. If the detailed balance condition is not satisfied, then the solution
$\rho_{mm}^{\rm ss}$ is a nonequilibrium steady-state. 

A useful connection to thermodynamics is provided if we consider that each state $m$
has a given energy $\epsilon_{m}$ and number of particles $N_{m}$ and that each 
different mechanism $\nu$ inducing transitions between states corresponds to  
a given reservoir with a given temperature $\beta^{-1}_{\nu}$ 
and chemical potential $\mu_{\nu}$.
We then assume that 
\begin{eqnarray}
\frac{\tilde{W}_{mm'}^{(\nu)}(0)}{\tilde{W}_{m'm}^{(\nu)}(0)} = 
\exp{\{-\beta_{\nu} (\epsilon_{m}-\epsilon_{m'}) + \beta_\nu\mu_{\nu} (N_{m}-N_{m'})  \}} .
\nonumber \\ \label{rateratio1res}
\end{eqnarray}
Equation~(\ref{detailbalance}) holds only if all the $\beta_{\nu}$s and $\mu_{\nu}$s are equal. 
In this case the equilibrium distribution correspond to the grand canonical ensemble.
If the $\beta_{\nu}$s and $\mu_{\nu}$s are different, the steady state is a
non-equilibrium steady state that obeys Eq.~(\ref{NESS}). 

%%%%%%%%%%%%%%%%%%%%%%%%%%%%%
\subsection{Long time behavior for $0< \alpha <1$}\label{longtimeGME2}

When $0< \alpha <1$, we must separately specify the statistical properties of the waiting time
distribution for the first jump after $t=0$.  We choose 
$\tilde{\phi}_{m}'(s)=\tilde{\phi}_{m}(s)$ because other choices add only further complications
but little of general interest to our specific problem.  
The inhomogeneous initial condition term then drops out, and for small $s$, 
using (\ref{ratebis}), Eq.~(\ref{GeneralizedMELaplace}) becomes
\begin{equation}
s \tilde{\rho}_m(s) - \rho_{m}(0) = \sum_{m',\nu} 
V_{mm'}^{(\nu)} s^{1-\alpha} \tilde{\rho}_{m'}(s). \label{GeneralizedMELaplaceAnomal}
\end{equation}
Using the rules of fractional calculus~\cite{Metzler}, this 
equation can be written in the time domain as 
\begin{equation}
\frac{d}{dt} \rho_m(t) = \sum_{m',\nu} V_{mm'}^{(\nu)} 
 \; _{0} D_{t}^{1-\alpha} \rho_{m'}(t) ,
\label{GeneralizedMEAnomal}
\end{equation}
where the Riemann-Liouville fractional integral is given by 
\begin{equation}
_{0} D_{t}^{1-\alpha} \rho_{m'}(t) =\frac{1}{\Gamma(\alpha)} \frac{d}{dt} 
\int_{0}^{t} d\tau (t-\tau)^{\alpha-1} \rho_{m'}(\tau) .
\label{RiemannLiouville}
\end{equation}
If the matrices with elements $V_{mm'}^{(\nu)}$ can be diagonalized and the eigenvalues
$v_i$ (which are all negative or zero) are non-degenerate,
the solution of Eq.~(\ref{GeneralizedMEAnomal}) can be written as a linear combination of 
solutions of 
\begin{equation}
\frac{d}{dt} \rho_{i}(t)= v_i \; _{0} D_{t}^{1-\alpha} \rho_{i}(t) . 
\label{GeneralizedMEAnomaldiag}
\end{equation}
The solution of this equation is the Mittag-Leffler function 
$E_{\alpha}(-(t/\tau)^{\alpha})$, where $\tau=(-v_i)^{1/\alpha}$
(see Appendix B of Ref.~\cite{Metzler}).  
At long times the Mittag-Leffler function decays as a power law,
\begin{equation}
E_{\alpha}(-(t/\tau)^{\alpha}) \sim \bigg((t/\tau)^{\alpha} \Gamma(1-\alpha) \bigg)^{-1}
\end{equation}
(at short times it behaves as a stretched exponential,
$E_{\alpha}(-(t/\tau)^{\alpha}) \sim 
\exp \bigg(-\frac{(t/\tau)^{\alpha}}{\Gamma(1+\alpha)} \bigg)$).
Thus the general solution of the generalized master equation is a linear 
combination of Mittag-Leffler functions, and the probability $\rho_m(t)$ 
decays toward the zero eigenvalue node as a power law. 

%%%%%%%%%%%%%%%%%%%%%%%%%%%%%%%%%%%%%%%%%%%%%%%%%%%%%%%%%%%%%%%%%%%%%%
\section{Counting statistics}\label{GFcurrentsec}

Although our ultimate goal is to establish conditions under which fluctuation theorems are valid for
systems whose dynamics are described by CTRWs, we first consider the counting statistics for
such a system.  These
statistics are interesting because they are experimentally accessible, and this analysis leads to
some definitions that are useful in the discussion of fluctuation theorems.

Consider a system described by the CTRW of Sec.~\ref{CTRW} where each 
microscopic state $m$ has a given number of particles $N_{m}$ and an energy 
$\epsilon_{m}$, and where the allowed transitions between pairs of states are 
due to different mechanisms $\nu$, each corresponding to an reservoir $\nu$. 
We want to calculate the probability $P(\{\Delta E^{(\nu)}\},\{\Delta N^{(\nu)}\},t)$
that an energy transfer $\Delta E^{(\nu)}$ and a matter transfer $\Delta N^{(\nu)}$
occurs between the system and the reservoir $\nu$ during time $t$.
We define the set of parameters $\boldsymbol{\gamma} \equiv (\{\gamma^{(\nu)}_{\rm e}\},
\{\gamma^{(\nu)}_{\rm m}\})$ and the generating function 
\begin{eqnarray}
\label{ProbTransferInverse}
\hspace{0cm}G(\i \boldsymbol{\gamma},t) &&= 
\int_{-\infty}^{\infty}\{d\Delta E^{(\nu)} \} \nonumber\\
&\times &\sum_{\{\Delta N^{(\nu)}\}=-\infty}^{\infty}
P(\{\Delta E^{(\nu)}\},\{\Delta N^{(\nu)}\},t)
\nonumber\\&\times&\hspace{0cm} \exp{\left[ \i \big( \gamma^{(\nu)}_{\rm e} \Delta E^{(\nu)}  
+ \gamma^{(\nu)}_{\rm m} \Delta N^{(\nu)} \big)\right]}.
\end{eqnarray}
The probability can be recovered from the generating function using
\begin{eqnarray}
\label{ProbTransfer}
P(\{\Delta E^{(\nu)}\},\{\Delta N^{(\nu)}\},t) &&= 
\int_{-\infty}^{\infty}\{\frac{d\gamma_{\rm e}^{(\nu)}}{2\pi}\} 
\int_{0}^{2\pi} \{\frac{d \gamma_{\rm m}^{(\nu)}}{2\pi}\} \nonumber \\
&&\hspace{-3.6cm} \times \exp{\{- \i \big( \Delta E^{(\nu)} 
\gamma^{(\nu)}_{\rm e} + \Delta N^{(\nu)} \gamma^{(\nu)}_{\rm m} \big)\}}
G(\i \boldsymbol{\gamma},t) \;.
\end{eqnarray}
Derivatives of the generating function with respect to the
elements of $\boldsymbol{\gamma}$ evaluated 
at $\boldsymbol{\gamma}=0$ gives the moments of the distribution of this process.\\

We define 
\begin{eqnarray}
\psi_{mm'}(\boldsymbol{\gamma},t) &\equiv&  \label{GFratemat} \\
&&\hspace{-0.8cm}\sum_{\nu} e^{\gamma^{(\nu)}_{\rm m} (N_{m}-N_{m'})}
e^{\gamma^{(\nu)}_{\rm e} (\epsilon_{m}-\epsilon_{m'})} \; \psi_{mm'}^{(\nu)}(t)
\nonumber \;,
\end{eqnarray}
which is the WTD matrix whose elements associated with a transition 
caused by mechanism $\nu$ are weighted by the exponential of $\gamma_{\rm e}^{(\nu)}$ 
($\gamma_{\rm m}^{(\nu)}$) times the change of energy (matter) that this transition induces.
We note that $\Psi(\boldsymbol{\gamma}=0,t)=\Psi(t)$.
To evaluate the generating function and the associated moments, we replace
the WTD $\psi_{mm'}^{(\nu)}(t)$ by (\ref{GFratemat}) 
in the CTRW (\ref{CTRW1}) and (\ref{CTRW2}), and thus obtain
\begin{eqnarray}
\label{CTRWGF1}
\eta_m(\boldsymbol{\gamma},t) &=& \sum_{m'\nu} 
\psi_{mm'}'^{(\nu)}(\boldsymbol{\gamma},t) \rho_{m'}(0) \\
&&+ \sum_{m'\nu} \int_{0}^{t} d\tau \psi_{m m'}^{(\nu)}
(\boldsymbol{\gamma},t-\tau) \eta_{m'}(\boldsymbol{\gamma},\tau) \nonumber
\end{eqnarray}
and 
\begin{eqnarray}
\label{CTRWGF2}
\rho_m(\boldsymbol{\gamma},t) = \phi_{m}'(t) \rho_{m}(0) 
+ \int_{0}^{t} d\tau \phi_{m}(t-\tau) \eta_{m}(\boldsymbol{\gamma},\tau) \;.
\end{eqnarray}
By doing so, we are weighting the probability of all the trajectories of length $t$ 
ending up in the state $m$ and along which a transfer of energy (matter)
between the system and the reservoir $\nu$ occurs, by the exponential of $\gamma_{\rm e}^{(\nu)}$ 
($\gamma_{\rm m}^{(\nu)}$) times this energy (matter) transfer.
By summing over all final states $m$, we reconstruct the generating
function~(\ref{ProbTransferInverse}) as
\begin{eqnarray}
G(\boldsymbol{\gamma},t) = \braket{I}{\rho(\boldsymbol{\gamma},t)}
= \sum_m \rho_m(\boldsymbol{\gamma},t) \;, \label{GFcurrent}
\end{eqnarray}
where $\ket{I}$ is the unit vector.
Proceeding as in Sec.~\ref{CTRW}, the formal solution of (\ref{CTRWGF1}) and
(\ref{CTRWGF2}) in Laplace space leads to the general solution for the generating function
\begin{eqnarray}
&&\hspace{-0.5cm}\tilde{G}(\boldsymbol{\gamma},s) \label{CTRWGFformalsol}\\ 
&&\hspace{0cm}= \bra{I} \bigg( \hat{\tilde{\Phi}}'(s) +\hat{\tilde{\Phi}}(s)  
\big(\hat{I}-\hat{\tilde{\Psi}}(\boldsymbol{\gamma},s) \big)^{-1} 
\hat{\tilde{\Psi}}'(\boldsymbol{\gamma},s) \bigg) \ket{\rho(0)} \;.\nonumber
\end{eqnarray}

If (\ref{CTRWGFformalsol}) can be evaluated and inverse Laplace transformed, it provides 
the full statistics of the energy and matter transfer for a finite time interval.
This is often a difficult task, and one therefore often focuses on the long-time behavior.
This behavior is accessed through the solution (\ref{CTRWGFformalsol}) and also
through the equation of
motion for the generating function, which can be deduced
by proceeding in the same way as in Sec.~\ref{GME} when deriving 
the generalized master equation from (\ref{CTRWGF1}) and (\ref{CTRWGF2}). 
Defining
\begin{equation}
\tilde{W}_{mm'}^{(\nu)}(\boldsymbol{\gamma},s) \equiv 
\frac{\tilde{\psi}_{mm'}^{(\nu)}(\boldsymbol{\gamma},s)}{\tilde{\phi}_{m'}(s)} \;,
\label{rateMod}
\end{equation}
we find 
\begin{eqnarray}
&&\dot{\rho}_m(\boldsymbol{\gamma},t) = \mathcal{I}(\boldsymbol{\gamma},t) 
+ \sum_{m',\nu} \int_{0}^{t} d\tau \label{GeneralizedMEMod} \\
&&\hspace{0cm} \times \bigg( W_{mm'}^{(\nu)}(\boldsymbol{\gamma},\tau) 
\rho_{m'}(\boldsymbol{\gamma},t-\tau)- W_{m'm}^{(\nu)}(\tau) 
\rho_{m}(\boldsymbol{\gamma},t-\tau) \bigg) \nonumber
\end{eqnarray}
where 
\begin{eqnarray}
\tilde{\mathcal{I}}(\boldsymbol{\gamma},s) &\equiv& \sum_{m',\nu} 
\bigg( \tilde{W}_{mm'}^{(\nu)}(\boldsymbol{\gamma},s) 
\big( \tilde{\phi}_{m'}(s) - \tilde{\phi}_{m'}'(s) \big) \rho_{m'}(0) \nonumber \\
&&\hspace{0.3cm}-  \tilde{W}_{m'm}^{(\nu)}(s) 
\big( \tilde{\phi}_{m}(s) - \tilde{\phi}_{m}'(s)\big) \rho_{m}(0) \bigg).
\label{inhomogMod}
\end{eqnarray}

To calculate the long-time behavior of the moments of the 
probability distribution of heat and 
matter transfer between the system and its reservoir,
we first consider the situation when a jump occurred at time zero.
We will comment later on situations in which this is not the case.
In the long time limit, (\ref{CTRWGFformalsol}) diverges at $\boldsymbol{\gamma}=0$,
because $\big(\hat{I}-\hat{\tilde{\Psi}}(s) \big)^{-1} 
\stackrel{s \to 0} = \big(\hat{I}-\hat{P} \big)^{-1}$,
a limit which is singular because the determinant of $\hat{I}-\hat{P}$ is zero. 
This can be seen by considering the transpose of the matrix (transposition does not affect 
the determinant) and by replacing its first column by the sum of all the columns of the 
matrix (the determinant is not affected by replacing a column by a linear combination of 
it with other columns), which only contains zeros since $\sum_{m'\neq m} P_{m'm}=1$ 
(the determinant of a matrix with a zero column is zero). 
To lowest order in $s$, the determinant behaves like $\sim s^{\alpha}$. 
This means that at small $s$, $\big(\hat{I}-\hat{\tilde{\Psi}}(s) \big)^{-1} 
\stackrel{s \to 0}{\sim} s^{-\alpha}$.
Using the long time behavior of the WTD as expressed in (\ref{s0WTDpsigeneral}) and 
(\ref{s0WTDphigeneral}), and using (\ref{CTRWGFformalsol}), we thus see that 
$\tilde{G}(\boldsymbol{\gamma}=0,s) \stackrel{s \to 0}{\sim} s^{-1}$. 
We could have arrived at this directly because $G(\boldsymbol{\gamma}=0,t)=1$
implies $\tilde{G}(\boldsymbol{\gamma}=0,s)=s^{-1}$. 
%However, when in the close neighborhood of $\boldsymbol{\gamma} =0$, at least one $\gamma$ is 
%nonzero, $\big(\hat{I}-\hat{\tilde{\Psi}}(\boldsymbol{\gamma},s) \big)^{-1}$ 
%is finite in the limit $s \to 0$, and we need to distinguish two cases.
%If $\alpha=1$, then $\lim_{s \to 0} \tilde{G}(\boldsymbol{\gamma},s)$ becomes 
%a constant, which means that $\lim_{t \to \infty} G(\boldsymbol{\gamma},t)=0$. 
%If $\alpha<1$, then $\tilde{G}(\boldsymbol{\gamma},s) \stackrel{s \to 0}{\sim} s^{\alpha-1}$
%which means (using Tauberian theorems~\cite{Hughes,Feller,Korevaar}) that 
%$G(\boldsymbol{\gamma},t) \stackrel{t \to \infty}{\sim} t^{-\alpha}$.
This reasoning makes it clear that in order to calculate moments, the derivatives 
with respect to one of the $\gamma$'s evaluated at $\boldsymbol{\gamma}=0$
must be calculated from~(\ref{CTRWGFformalsol}) before the long time limit is taken.

Upon taking the $n$th derivative of~(\ref{CTRWGFformalsol}) with respect to one of 
the $\gamma$'s at $\boldsymbol{\gamma}=0$, the dominant contribution at small $s$ is
\begin{eqnarray}
&&\hspace{-0.2cm}\partial_{\gamma}^{n} \tilde{G}(\boldsymbol{\gamma}=0,s) \label{Nouveau}\\
&&\hspace{-0.2cm}\stackrel{s \to 0}{=} \bra{I} \hat{B} s^{\alpha-1} 
\big( \partial_{\gamma}^{n} \big(\hat{I}-\hat{\tilde{\Psi}}
(\boldsymbol{\gamma}=0,s) \big)^{-1} \big)
\hat{P}(\boldsymbol{\gamma}=0) \ket{\rho(0)} \nonumber \\
&&\hspace{-0cm} \sim s^{-n \alpha-1} \;. \nonumber
\end{eqnarray}
We used the fact that at small $s$,
$\partial_{\gamma}^{n} \big(\hat{I}-\hat{\tilde{\Psi}}(\boldsymbol{\gamma}=0,s) 
\big)^{-1}$ $\sim s^{-\alpha (n+1)}$.
$\hat{B}$ is a diagonal matrix with diagonal elements $B_m$.
%If $\hat{M}(\gamma)$ is an arbitrary matrix depending on $\gamma$, 
%$\partial_{\gamma} \hat{M}^{-1}(\gamma) = \hat{M}^{-1}(\gamma) 
%\big( \partial_{\gamma} \hat{M}(\gamma) \big) \hat{M}^{-1}(\gamma)$.
%We can use this relation to express $\partial_{\gamma}^n \hat{M}^{-1}(\gamma)$
%as a sum of different terms containing only $\hat{M}^{-1}(\gamma)$'s and 
%$\partial_{\gamma} \hat{M}(\gamma)$'s.
%If $\hat{M}(\gamma)=\hat{I}-\hat{\tilde{\Psi}}(\gamma,s)$, when $s \to 0$,
%$\partial_{\gamma} \hat{M}(\gamma)$ is finite but $\hat{M}^{-1}(\gamma)$ 
%diverge as $s^{-1}$. Therefore, the dominant terms in 
%$\partial_{\gamma}^n \hat{M}^{-1}(\gamma)$ are the ones 
%with the most $\hat{M}^{-1}(\gamma)$'s terms, which is $n+1$.
%These terms therefore behave as $s^{-\alpha (n+1)}$ which concludes our proof. 
%We understand now that (\ref{Nouveau}) implies 
%\begin{eqnarray}
%&&\hspace{-0.5cm} \partial_{\gamma}^{n} \tilde{G}(\boldsymbol{\gamma}=0,s)
%\sim s^{-n \alpha-1} \;. 
%\end{eqnarray}
Using Tauberian theorems~\cite{Hughes,Feller,Korevaar}, we conclude that 
the moments of the energy and matter transfer between the system and the 
reservoir $\nu$ will behave at long times as 
\begin{eqnarray}
\mean{(\Delta E^{(\nu)})^n} \; , \; \mean{(\Delta N^{(\nu)})^n} 
\stackrel{t \to \infty}{\sim} t^{n \alpha} .
\label{momentasymptoticEnergymatter}
\end{eqnarray}
As noted earlier, these moments are experimentally accessible.
%REFERENCE?

Let us now turn back to the case where no jump occurred at time zero.
We only comment on WTDs with $\alpha = 1$ since only then is it possible to carry
out the averaging procedure described in Sec.~\ref{CTRW}. 
Since $\tilde{\Psi}_{mm'}'(s) \stackrel{s \to 0}{=} t_{mm'}/t_{m'}$ 
and $\tilde{\Psi}_{mm'}(s) \stackrel{s \to 0}{=}P_{mm'}$, 
Eq.~(\ref{momentasymptoticEnergymatter}) with $\alpha=1$ is still valid, 
perhaps with a different proportionality factor. 

In Appendix~\ref{QD} we explicitly implement these ideas by calculating
the moments for a two level quantum dot.

%%%%%%%%%%%%%%%%%%%%%%%%%%%%%%%%%%%%%%%%%%%%%%%%%%%%%%%%%%%%%%%%%%%%%%
\section{Fluctuation theorems}\label{Ptraj}

We now explore the conditions under which a fluctuation theorem holds for a CTRW.
Two types of derivations have been used to obtain steady state 
fluctuation theorems for stochastic dynamics. 
The first relies on a symmetry of the generating function,
which translates into a fluctuation theorem using large deviation theory 
as in Refs.~\cite{Lebowitz99,AndrieuxGaspard07a}. The other exploits
the specific form of the logarithm of the ratio of the probability 
of a trajectory and the probability of its time reversed trajectory.   
We will use both for CTRWs and find that they can 
only be considered equivalent under specific conditions. 

%%%%%%%%%%%%%%%%%%%%%%%%%%%%%%%%%%%%%%%%%%%%%%%%%%%%%%%%%%%%%%%%%%%%%%
\subsection{Using large deviation} \label{SSFlucTheo}

This approach to arrive at a fluctuation theorem can only be implemented for $\alpha=1$ because it
requires a finite mean waiting time between transitions.  We thus restrict our discussion to this
case.  We have seen in Sec.~\ref{longtimeGME} that as long as one considers WTDs with $\alpha=1$,
the generalized master equation at long times behaves like a Markovian master equation and thus
reaches a steady state. Using the same arguments, we can show that
the equation of motion~(\ref{GeneralizedMEMod}) for the generating function 
behaves at long times as
\begin{eqnarray}
\frac{\partial}{\partial t} \ket{G(\boldsymbol{\gamma},t)} \underset{t \to \infty}{=} 
\boldsymbol{W}(\boldsymbol{\gamma}) \ket{G(\boldsymbol{\gamma},t)} \;,
\label{eqmotionGFalphaone}
\end{eqnarray}
where
\begin{eqnarray}
&&[\boldsymbol{W}(\boldsymbol{\gamma})]_{mm'} 
\equiv \tilde{W}_{mm'}(\boldsymbol{\gamma},0) \ \ {\rm for} \ \ m \neq m' \nonumber\\
&&[\boldsymbol{W}(\boldsymbol{\gamma})]_{mm}
\equiv - \sum_{n} \tilde{W}_{nm}(0)  . \label{DefrateGF}
\end{eqnarray}
This implies that for long times
\begin{eqnarray}
G(\boldsymbol{\gamma},t) = C(\boldsymbol{\gamma},t) \e^{S(\boldsymbol{\gamma}) t} , 
\end{eqnarray}
where $\lim_{t \to \infty} \frac{1}{t} \ln C(\boldsymbol{\gamma},t)=0$ 
and where $S(\boldsymbol{\gamma})$ is the dominant eigenvalue of 
$\tilde{\boldsymbol{W}}(\boldsymbol{\gamma},0)$. 
This dominant eigenvalue gives the cumulant generating function because
\begin{eqnarray}
S(\boldsymbol{\gamma}) = \lim_{t \to \infty} \frac{1}{t} \ln G(\boldsymbol{\gamma},t) \;.
\label{cumulantGFlong}
\end{eqnarray}
Note that the derivation to follow does not explicitly require the generalized master equation to be
equivalent to a Markovian master equation at long times; what is required is the limiting
behavior~(\ref{cumulantGFlong}). Furthermore,
using~(\ref{rateratio1res}), we can verify that (\ref{DefrateGF}) satisfies 
\begin{eqnarray}
\boldsymbol{W}(\boldsymbol{\gamma}) =
\boldsymbol{W}^{t}(\boldsymbol{A}-\boldsymbol{\gamma}) \;,
\label{symmonWflux}
\end{eqnarray}
where $\boldsymbol{A}=(\{\beta_{\nu}\},\{-\beta_{\nu} \mu_{\nu}\})$.
Since these two matrices have the same eigenvalues, this implies the symmetry
\begin{eqnarray}
S(\boldsymbol{\gamma}) = S(\boldsymbol{A}-\boldsymbol{\gamma}) \;.
\label{symmetryVP}
\end{eqnarray}
%%%%%%%%%%%%%%%%%%%%%%%%%
%%%%%%%%%%%%%%%%%%%%%%%%%
%%%%%%%%%%%%%%%%%%%%%%%%%
%%%%%%%%%%%%%%%%%%%%%%%%%
Large deviation theory~\cite{Lebowitz99,AndrieuxGaspard07a} can now be applied.
The limiting behavior~(\ref{cumulantGFlong}) and the symmetry~(\ref{symmetryVP}) then imply
the fluctuation theorem for the probability $P\bigg(\{-\frac{1}{t} \Delta E^{(\nu)}\},\{-\frac{1}{t}
\Delta N^{(\nu)}\} \bigg)$ for the energy and matter currents (cf. below) between the system and the
reservoir $\nu$ at long times~\cite{Lebowitz99,AndrieuxGaspard07a},
%\begin{eqnarray}
%&&\hspace{-0.5cm}\frac{P\bigg(\{\frac{1}{t} \int_{0}^{t} d\tau I^{(\nu)}_{\rm e}(\tau)\},
%\{\frac{1}{t} \int_{0}^{t} d\tau I^{(\nu)}_{\rm m}(\tau)\} \bigg)}
%{P\bigg(\{-\frac{1}{t} \int_{0}^{t} d\tau I^{(\nu)}_{\rm e}(\tau)\},
%\{-\frac{1}{t} \int_{0}^{t} d\tau I^{(\nu)}_{\rm m}(\tau)\} \bigg)} 
%= \e^{\Delta S_r(t)} \nonumber\\ &&\hspace{5cm}
%\ \ {\rm for} \ \ t \to \infty \label{detailedFTCurrent} \;,
%\end{eqnarray}
\begin{eqnarray}
&&\hspace{-0.5cm}\frac{P\bigg(\{\frac{1}{t} \Delta E^{(\nu)}\},
\{\frac{1}{t} \Delta N^{(\nu)} \} \bigg)}{P\bigg(\{-\frac{1}{t} \Delta E^{(\nu)}\},
\{-\frac{1}{t} \Delta N^{(\nu)}\} \bigg)} 
= \e^{\Delta S_r} 
\ \ {\rm for} \ \ t \to \infty . \nonumber \\ \label{detailedFTCurrent} 
\end{eqnarray}
Here
%\begin{eqnarray}
%\Delta S_r = \sum_{\nu} \bigg( 
%\beta_{\nu} \int_{0}^{t} d\tau I^{(\nu)}_{\rm e}(\tau) 
%- \beta_{\nu} \mu_{\nu} \int_{0}^{t} d\tau I^{(\nu)}_{\rm m}(\tau) \bigg) \;.
%\nonumber\\ \label{FTcurrents}
%\end{eqnarray}
\begin{eqnarray}
\Delta S_r = \sum_{\nu} \bigg( \beta_{\nu} \Delta E^{(\nu)} 
- \beta_{\nu} \mu_{\nu} \Delta N^{(\nu)} \bigg) \label{FTcurrents}
\end{eqnarray}
represents the change of entropy due to the exchange processes with the reservoirs. 
We note that the change in energy and matter can be written 
in terms of energy and matter currents as 
\begin{eqnarray}
\label{Ec}
\Delta E^{(\nu)} &\equiv& \int_{0}^{t} d\tau I_{\rm e}^{(\nu)}(\tau) \\
\Delta N^{(\nu)} &\equiv& \int_{0}^{t} d\tau I_{\rm m}^{(\nu)}(\tau) \;,
\label{Nc}
\end{eqnarray}
so that in the long time limit $\frac{1}{t} \Delta E^{(\nu)}$ and 
$\frac{1}{t} \Delta N^{(\nu)}$ correspond to steady state currents.
This is why (\ref{detailedFTCurrent}) is called a current 
fluctuation theorem \cite{AndrieuxGaspard07a}.

We now define the matrix $\bar{\boldsymbol{W}}(\bar{\gamma})$ as 
$\boldsymbol{W}(\boldsymbol{\gamma})$ where $\boldsymbol{\gamma}$ 
is replaced by $\bar{\boldsymbol{\gamma}}=(\{\gamma_{\rm e} \beta_{\nu} \},
\{-\gamma_{\rm m} \beta_{\nu} \mu_{\nu} \})$.
Obviously, when replacing $\boldsymbol{W}(\boldsymbol{\gamma})$
by $\bar{\boldsymbol{W}}(\bar{\gamma})$ in our previous results,
we calculate the statistics of $\Delta S_r$.
The symmetry (\ref{symmonWflux}) now implies that
$\bar{\boldsymbol{W}}(\bar{\gamma}) = \bar{\boldsymbol{W}}^{t}(1-\bar{\gamma})$ 
so that
\begin{eqnarray}
\bar{S}(\bar{\gamma}) = \bar{S}(1-\bar{\gamma}) \;.
\label{symmonSentropy}
\end{eqnarray}
Using again large deviation theory, we get the fluctuation theorem
\begin{eqnarray}
\frac{P\bigg(\frac{1}{t} \Delta S_r\bigg)}{P\bigg(-\frac{1}{t} \Delta S_r\bigg)} 
= \e^{\Delta S_r} \ \ {\rm for} \ \ t \to \infty . \label{detailedFTres}
\end{eqnarray}
In the steady state, the average energy and matter transferred with a reservoir 
$\nu$, can be obtained by taking the derivative with respect to $\gamma_{\nu}$ at 
$\boldsymbol{\gamma}=0$ of the formal solution of (\ref{eqmotionGFalphaone}).
We get 
\begin{eqnarray}
\label{currentaverages}
&&\hspace{-0.4cm}\mean{\Delta E^{(\nu)}}/t \equiv \mean{I^{(\nu)}_{\rm e}} = 
\sum_{m,m'} (\epsilon_m-\epsilon_{m'}) W_{mm'}^{(\nu)} \rho^{\rm ss}_{m'} \\
&&\hspace{-0.4cm}\mean{\Delta N^{(\nu)}}/t \equiv \mean{I^{(\nu)}_{\rm m}} 
= \sum_{m,m'} (N_m-N_{m'}) W_{mm'}^{(\nu)} \rho^{\rm ss}_{m'} \;.\nonumber
\end{eqnarray}
Current conservation at steady state follows from
\begin{eqnarray}
\sum_{\nu} \mean{I^{(\nu)}_{\rm e}} = \sum_{\nu} \mean{I^{(\nu)}_{\rm m}}=0
\label{currentconservation} .
\end{eqnarray}
If we assume that all $M$ reservoirs ($\nu=1,\hdots,M$) have different 
temperatures and chemical potentials, we have $M-1$ independent 
nonequilibrium forces associated to energy and matter transfer  
that can be defined as
\begin{eqnarray}
X_{\rm e}^{(i)} = \beta_1-\beta_{i+1} \ \ , \ \ 
X_m^{(i)} = - \beta_1 \mu_1 + \beta_{i+1} \mu_{i+1} ,
\end{eqnarray}
where $i=1,\hdots,M-1$.
Using (\ref{currentconservation}), the average change in the entropy
due to exchange processes with the reservoirs $\mean{\Delta S_r}$ can 
be written in the familiar thermodynamical form of the entropy 
production in a steady state   
\begin{eqnarray}
\mean{\Delta S_r}/t = \sum_{i=1}^{M-1} \big( X_{\rm e}^{(i)} 
\mean{I^{(i)}_{\rm m}} + X_{\rm m}^{(i)} \mean{I^{(i)}_{\rm m}} \big) .
\label{EntropyProdThermo}
\end{eqnarray}

When $0< \alpha <1$, we have seen in section \ref{longtimeGME2} that the 
solution of the GME behaves at long times as a power law $t^{- \alpha}$. 
This will also be the case for the solution of the equation of 
motion for the generating function (\ref{GeneralizedMEMod}).
Therefore, contrary to (\ref{cumulantGFlong}), the cumulant generating function 
$\lim_{t \to \infty} \frac{1}{t} \ln G(\boldsymbol{\gamma},t)$ is zero.
This indicates that the cumulants decay slower than $t$. 
The fluctuation theorem symmetry \label{symmertryVP} is only present in the 
eigenvalues of the generator of the GME (\ref{eqmotionGFalphaone}) but not 
in the eigenvectors. 
For $\alpha=1$, the cumulant generating function is given by the dominant 
eigenvalue, so that the fluctuation theorem will reflect itself on all 
quantities relates to it (large deviation theory is precisely used to 
make the link between this generating function and the probabilities). 
However, in the case $0< \alpha <1$, there is no way to separate in the long 
time limit, the contribution of the eigenvalues from the contribution of the 
eigenvectors to the statistics thus preventing a fluctuation theorem to hold.

%%%%%%%%%%%%%%%%%%%%%%%%%%%%%%%%%%%%%%%%%%%%%%%%%%%%%%%%%%%%%%%%%%%%%%
\subsection{Using time reversal symmetry}\label{TRev}
 
\begin{figure}[h]
\centering
\rotatebox{0}{\scalebox{0.5}{\includegraphics{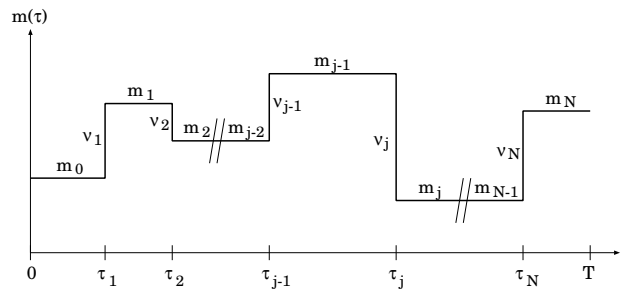}}}
\caption{Representation of a trajectory $m_{(\tau)}$.}
\label{figTraj}
\end{figure}

We denote a forward trajectory of the system between times $t=0$ and $t=T$ by $m_{\tau}$.  
As illustrated in Fig.~\ref{figTraj}, at $t=0$ the system is in state $m_0$ and stays
there until it jumps to state $m_1$ at time $\tau_1$ via mechanism $\nu_1$.  
It remains there until time $\tau_2$, when it jumps to state $m_2$ via mechanism $\nu_2$.  
The trajectory continues in this fashion; at time $\tau_N$ there is a jump to state $m_N$, 
where the system remains at least until time $T$. 
The total number of jumps in this trajectory is $N$.
The probability of this trajectory is 
\begin{eqnarray}
\label{forwardProb}
&&P[m_{\tau}]=\rho_{m_0}(0) \psi_{m_1 m_0}'^{(\nu_1)}(\tau_1) \nonumber \\
&&\hspace{1.5cm}\times \bigg( \prod_{i=1}^{N-1} \psi_{m_{i+1}m_{i}}^{(\nu_{i+1})}
(\tau_{i+1}-\tau_{i}) \bigg) \phi_{m_{N}}(T-\tau_N). \nonumber \\
&&
\end{eqnarray}

%%%%%%%%%%%%%%%%%%%%%%%%%
%%%%%%%%%%%%%%%%%%%%%%%%%

The time-reversed trajectory $\bar{m}_{\tau}$ starts in state $m_N$ at time $T$, jumps to state
$m_{N-1}$ at time $T-\tau_N$ via mechanism $\nu_N$, and so on.  At time $T-\tau_1$ a jump  
to state $m_1$ occurs via mechanism $\nu_1$, and the system remains there until at least time $0$.
The probability of this time reversed trajectory is 
\begin{eqnarray}
\label{backwardProb}
&&P[\bar{m}_{\tau}] = \rho_{m_N}(T) \psi_{m_{N-1} m_{N}}'^{(\nu_{N-1})}(T-\tau_N) \nonumber \\
&&\hspace{1.5cm}\times \bigg( \prod_{i=0}^{N-2} \psi_{m_{i}m_{i+1}}^{(\nu_{i})}
(\tau_{i+2}-\tau_{i+1}) \bigg) \phi_{m_{0}}(\tau_1). \nonumber \\
&&
\end{eqnarray}

Next we consider the quantity
\begin{equation}
%\label{ratioprob}
R[m_{\tau}] \equiv \ln \frac{P[m_{\tau}]}{P[\bar{m}_{\tau}]} ,
\label{entropyprod}
\end{equation}
whose explicit form reads
\begin{eqnarray}
R[m_{\tau}] &=& \ln \frac{\rho_{m_0}(0)}{\rho_{m_N}(T)} 
+ \ln \prod_{i=1}^{N-1} \frac{\psi_{m_{i+1}m_{i}}^{(\nu_{i+1})}(\tau_{i+1}-\tau_{i})}
{\psi_{m_{i-1}m_{i}}^{(\nu_{i})}(\tau_{i+1}-\tau_{i})} \nonumber\\
&&\hspace{0cm} +\ln \frac{\psi_{m_1 m_0}'^{(\nu_1)}(\tau_1)  
\phi_{m_{N}}(T-\tau_N)}{\psi_{m_{N-1} m_{N}}'^{(\nu_{N})}(T-\tau_N) 
\phi_{m_{0}}(\tau_1)} \;. \label{Rgeneral} 
\end{eqnarray}
Because each $\bar{m}_{\tau}$ is the unique mirror image of $m_{\tau}$, a
sum over all possible forward trajectories is equivalent to a sum over all 
possible time-reversed trajectories.
Therefore, since normalization implies that $\sum_{\bar{m}_{\tau}} P[\bar{m}_{\tau}]=1$, an 
integral fluctuation theorem follows immediately from the definition~(\ref{entropyprod}), 
\begin{equation}
\mean{\e^{-R}}= \sum_{m_{\tau}} \e^{-R[m_{\tau}]} P[m_{\tau}] = 1 .
\label{integralFT}
\end{equation}
The positivity of the ensemble average of $R[m_{\tau}]$,
$\mean{R} \geq 0$, then follows from Jensen's inequality. 
Using the important property $R[m_{\tau}]=-R[\bar{m}_{\tau}]$, which follows form 
(\ref{entropyprod}) together with the fact that by taking twice the time-reversal of a 
trajectory we get back to the original trajectory $\bar{\bar{m}}_{\tau}=m_{\tau}$,
we can also derive a detailed fluctuation theorem for $R$,
\begin{eqnarray}
P(R) &=& \sum_{m_{\tau}} \delta(R-R[m_{\tau}]) P[m_{\tau}] \nonumber \\
&=& \sum_{m_{\tau}} \delta(R-R[m_{\tau}]) \e^{R[m_{\tau}]} P[\bar{m}_{\tau}] \nonumber \\
&=& \e^{R} \sum_{\bar{m}_{\tau}} \delta(R-R[m_{\tau}]) P[\bar{m}_{\tau}] \nonumber \\
&=& \e^{R} \sum_{\bar{m}_{\tau}} \delta(R+R[\bar{m}_{\tau}]) P[\bar{m}_{\tau}] \nonumber \\
&=& \e^{R} P(-R) \label{detailedFT} \;.
\end{eqnarray}
The integral fluctuation theorem (\ref{integralFT}) and the detailed 
fluctuation theorem (\ref{detailedFT}) for R are thus completely general 
and valid for any continuous time random walk.
It is worth mentioning that a similar derivation can be done for any 
dynamics as long as each trajectory has a corresponding time-reversed 
trajectory with a nonzero probability \cite{CleurenVDBroeckKawai}.

%In~\cite{QianWang} it is argued that a necessary condition for a CTRW to satisfy
%microreversibility is the separability of the waiting time distribution into the
%product of a waiting time portion that depends only on the originating state,
%and a transition matrix that connects a given initial and final state:
%\begin{equation}
%\psi_{mm'}^{(\nu)}(t) = P_{mm'}^{(\nu)} \; \psi_{m'}(t).
%\label{DTI}
%\end{equation}
%We note that it is implicitly assumed that the WTDs have a finite first moment (see below).
%Another necessary condition is that the system also obeys detailed balance (which
%it cannot do if different transition mechanisms have different statistics, see below).
%Together, these two necessary condition constitute a sufficient condition.
%(The separability condition is called ``direction time independence"
%by Qian and Wang~\cite{QianWang,WangQian}.)
%We will see that separability also plays a role in the derivation of a fluctuation theorem.

To make these fluctuation theorems useful, we need to 
give a physical interpretation to $R[m_{\tau}]$.
In particular, we will argue that $R$ can only be interpreted as a change of entropy
if two conditions are satisfied.  One is that the
WTDs have a finite first moment ($\alpha=1$).  The other is that they be
separable~\cite{WangQian,QianWang}, that is, that the waiting time distributions can be written as
the product of a waiting time portion that depends only on the originating state, and a transition
matrix that connects given initial and final states,
\begin{equation}
\psi_{mm'}^{(\nu)}(t) = P_{mm'}^{(\nu)} \; \psi_{m'}(t)
\label{DTI}
\end{equation} 
the separability condition is called ``direction time independence" 
by Qian and Wang~\cite{WangQian,QianWang}).
The first term of Eq.~(\ref{Rgeneral}),
\begin{equation}
\Delta S[m_{\tau}] \equiv \ln \frac{\rho_{m_0}(0)}{\rho_{m_N}(T)}
= S_{m_N}(T) - S_{m_0}(0) ,
\label{TrajentropyGibbs}
\end{equation}
where $S_{m}(t)=-\ln \rho_{m}(t)$, 
can be interpreted as a change of system (Gibbs) entropy along the trajectory because it depends
only on the initial and final microscopic states of the system for that trajectory, and the
average over trajectories is then simply $\mean{\Delta S}= S(t)-S(0)$, where
$S(t)=\sum_m \rho_{m}(t) S_{m}(t)$ is just a straightforward average over states.  

In order to interpret the two remaining terms in (\ref{Rgeneral}), 
we implement separability of the WTDs so that  
\begin{eqnarray}
R[m_{\tau}] &=& \ln \frac{\rho_{m_0}(0)}{\rho_{m_N}(T)} 
+\ln \bigg( \prod_{i=0}^{N-1} \frac{P_{m_{i+1}m_{i}}^{(\nu_{i+1})}}
{P_{m_{i}m_{i+1}}^{(\nu_{i+1})}} \bigg) \nonumber\\
&&+\ln \frac{\psi_{m_0}'(\tau_1) \phi_{m_{N}}(T-\tau_N)} 
{\psi_{m_{N}}'(T-\tau_N) \phi_{m_{0}}(\tau_1)} \;. \label{ratioprobexpli} 
\end{eqnarray}
Now only the WTDs of the first jumps remain.
In Eq.~(\ref{detailedFT}), the path summation runs over all possible 
trajectories including those with or without a jump at time zero. 
%Even if one would only considers forward trajectories without a 
%jump at time zero, the corresponding time-revered trajectories 
%would sometimes have a jump at time zero and sometimes not.
One way to handle the problem of having to treat the first jump differently 
from the others is via the time averaging procedure (\ref{WTfirstjump}) 
in (\ref{forwardProb}) as well as in (\ref{backwardProb}). 
This can only be done if the first moments are finite.
The third term in (\ref{ratioprobexpli}) now becomes 
equal to $\ln (t_{m_N}/t_{m_0})$, so that 
using (\ref{SumrateMarkov}) the second and third terms in (\ref{ratioprobexpli}) 
can be combined and $R[m_\tau]$ can be written in terms of the transition matrix elements as
\begin{eqnarray}
R[m_{\tau}] = \ln \frac{\rho_{m_0}(0)}{\rho_{m_N}(T)} 
+\ln \bigg( \prod_{i=0}^{N-1} \frac{\tilde{W}_{m_{i+1}m_{i}}^{(\nu_{i+1})}(0)}
{\tilde{W}_{m_{i}m_{i+1}}^{(\nu_{i+1})}(0)} \bigg) \;. \label{formestandard}
\end{eqnarray}
This form of $R[m_{\tau}]$ is now exactly the same as the one derived 
for the Markovian master equation (\ref{MEfake}) (see~\cite{Seifert05,EspositoHarbola07}). 
This is a manifestation of the so-called corresponding 
Markov process of a CTRW~\cite{QianWang}. 

We denote the second term on the right hand side of (\ref{formestandard}) 
as $\Delta S_r[m_{\tau}]$ and call it the reservoir part of the trajectory 
entropy because it can be interpreted as the change in entropy along the 
trajectory due to exchange processes between the system and the reservoirs. 
Indeed, using (\ref{rateratio1res}), this term can be expressed as
\begin{eqnarray}
\Delta S_r[m_{\tau}] &\equiv& \ln \bigg( \prod_{i=0}^{N-1} 
\frac{\tilde{W}_{m_{i+1}m_{i}}^{(\nu_{i+1})}(0)}
{\tilde{W}_{m_{i}m_{i+1}}^{(\nu_{i+1})}(0)} \bigg) \nonumber\\
&&\hspace{-1.5cm}=\sum_{\nu} \bigg( -\beta_{\nu} \Delta E^{(\nu)}[m_{\tau}]
+ \beta_{\nu} \mu_{\nu} \Delta N^{(\nu)}[m_{\tau}] \bigg) ,
\label{to_be_labeled}
\end{eqnarray}
where the change of energy and number of particles along the trajectory 
due to the mechanism $\nu$ can be expressed in terms of heat and matter 
currents along the trajectory as
\begin{eqnarray}
\Delta E^{(\nu)}[m_{\tau}] &\equiv& \int_{0}^{t} d\tau I_e^{(\nu)}[m_{\tau}] \\
\Delta N^{(\nu)}[m_{\tau}] &\equiv& \int_{0}^{t} d\tau I_m^{(\nu)}[m_{\tau}]
\end{eqnarray}
[cf. Eqs.~(\ref{Ec}) and (\ref{Nc})].
These currents are sequences of $\delta$ functions centered at the times of 
the jumps and multiplied by the corresponding energy or matter change.
They are positive (negative) if energy or matter increases (decreses) in the system.

We have thus arrived at the important result that for separable WTDs with $\alpha=1$, 
$R[m_{\tau}]=\Delta S[m_{\tau}]+\Delta S_r[m_{\tau}]$. Since
$\Delta S[m_{\tau}]$ and $\Delta S_r[m_{\tau}]$ are interpreted respectively
as the change in the system entropy along the trajectory and the change in entropy due to
the exchange processes between the system and its reservoirs, it is natural 
to interpret $R[m_{\tau}]$ as the total change in entropy along the 
trajectory (also called the total change in the trajectory entropy 
production)~\cite{Seifert05,EspositoHarbola07}.
Note that in this case the principle of microreversibility, implying that 
at equilibrium the probability of a forward trajectory is identical to the 
probability of its time reversed trajectory [$R[m_{\tau}]=0$], is satisfied 
if the detailed balance condition (\ref{detailbalance}) is satisfied and if 
the system is initially at equilibrium [so that 
$\rho_{m_0}(0)=\rho_{m_0}^{\rm eq}$ and $\rho_{m_N}(T)=\rho_{m_T}^{\rm eq}$].
This result is consistent with the findings of Ref.~\cite{QianWang} 
stating that separability of the WTDs and detailed balance are 
sufficient conditions for microreversibility to be satisfied in a CTRW.

In Sec.~\ref{SSFlucTheo} we showed that a fluctuation theorem for 
$\Delta S_r$ can be derived for long times [see Eq.~(\ref{detailedFTres})].
In fact, considering separable WTDs with $\alpha=1$, the fluctuation 
theorem (\ref{detailedFTres}) can be seen as resulting from the 
fluctuation theorem (\ref{detailedFT}), as follows. 
The quantity $\Delta S[m_{\tau}]$ is a bounded quantity which only depends 
on the probability distribution of the initial and final states of the trajectory. 
On the contrary, $\Delta S_r[m_{\tau}]$ changes each 
time a jump occurs along the system trajectory.
It is therefore reasonable to assume that for very long trajectories 
and for the huge majority of realizations, the contribution from 
$\Delta S_r[m_{\tau}]$ in $R[m_{\tau}]=\Delta S[m_{\tau}]
+\Delta S_r[m_{\tau}]$ will be significantly dominant so that in the 
long time limit (\ref{detailedFT}) reduces to (\ref{detailedFTres}).
However, the derivation of (\ref{detailedFTres}) only required WTDs 
with $\alpha=1$, but did not require separability. 
The reason for this requirement in order to identify Eq.~(\ref{detailedFT})
as a useful fluctuation theorem for all times is that otherwise
$R[m_{\tau}]$ has no clear physical interpretation.
The problem is that $R[m_{\tau}]$ (and, more specifically, the second term 
in (\ref{Rgeneral}), which according to our previous argument dominates at 
long times) depends on the time intervals between the jumps along the trajectory. 
This implies that at the level of a single trajectory, $R[m_{\tau}]$ 
cannot be expressed in terms of exchange processes with the 
reservoirs (more precisely, in terms of time integrated currents).
However, the fluctuation theorem (\ref{detailedFTres}) indicates that at long 
times the probability $P(R)$ to observe a trajectory such that $R[m_{\tau}]=R$ 
becomes equivalent to the probability to observe a trajectory 
with a set of time integrated currents so that, via (\ref{FTcurrents}), $\Delta S_r \approx R$.
One way to understand this result is to coarse-grain the trajectories in time.
Instead of specifying exactly the time at which each jump occurs, we define 
small time intervals of equal size $dt$ (sufficiently small so that the probability 
of observing two transitions in one is interval negligible, i.e., small compared to 
the mean time for a transition to occur).  We then specify whether a transition between 
two states occurred or not in this interval.
In this way, we define coarse-grained trajectories (denoted by $\tilde{m}_{\tau}$),
and we note that different microscopic trajectories can lead to the same coarse 
grained trajectory.
To calculate the probabilities of these trajectories, we use the fact that at long 
times the dynamics is described by (\ref{MEfake}), which can be discretized in time intervals $dt$.
This discretized form allows us to identify the probability to stay in a given state 
$m$ during a given time interval with $1-\sum_{m',\nu} \tilde{W}_{m'm}^{\nu}(0) dt$, and the 
probability to jump from $m$ to $m'$ by mechanism $\nu$ with $\tilde{W}_{mm'}^{\nu}(0) dt$.
Using these probabilities, we can construct the probability of a coarse-grained 
trajectory and that of its time reversed coarse-grained trajectory.
The logarithm of their ratio gives $\Delta S[\tilde{m}_{\tau}]+\Delta S_r[\tilde{m}_{\tau}]$, 
with the same definitions as in (\ref{TrajentropyGibbs}) and (\ref{to_be_labeled}), 
but for $\tilde{m}_{\tau}$ instead of $m_{\tau}$. 
With the argument that at long times $\Delta S_r[\tilde{m}_{\tau}]$ will be dominant,
we understand why the fluctuation theorem (\ref{detailedFTres}) holds at long times.
One should note that long times are also needed in this case to get rid of the 
contribution from the initial part of the trajectory where (\ref{MEfake}) is not valid. 
Also, contrary to the case where separability holds, it is only 
at the coarse grained level and considering very long trajectories 
that such a dynamics can satisfy "microscopic" reversibility.
The fluctuation theorem (\ref{detailedFTres}) does not result from a probabilistic 
asymmetry under time reversal at the microscopic trajectory level as in the case of 
separable WTDs, but at a coarse grained level and only for long times.

We can finally comment on the case of WTDs with diverging mean waiting times. 
Even if separability is satisfied, because mean waiting times diverge, 
(\ref{rateratio1res}) cannot be used to express $R[m_{\tau}]$ 
(or parts of it) in term of exchange processes with the reservoirs.
The fluctuation theorem (\ref{detailedFT}) is thus valid but seemingly useless.

%%%%%%%%%%%%%%%%%%%%%%%%%%%%%%%%%%%%%%%%%%%%%%%%%%%%%%%%%%%%%%%%%%%%%%
\section{Conclusions}\label{conc}

In this paper we have considered continuous time random walks (CTRWs) in which 
multiple mechanisms that may have different statistical properties induce
transitions between pairs of states.
A given energy and number of particles can be attributed to each state,
so that each mechanism can be thought of as corresponding to a given reservoir.
If these reservoirs have different statistical properties,
such a CTRW therefore describes the stochastic dynamics of an open system.
The statistics of the transitions between states associated with 
each mechanism are described by waiting time distributions (WTDs).
If a WTD decays at long times at least as fast as $t^{-2}$, the 
distribution has a finite first moment and we say that $\alpha=1$.
If the decay is slower, as $t^{-\alpha-1}$ where $0<\alpha<1$, all moments diverge.

We have analyzed the long-time behavior of the probability $\rho_m(t)$ that the system 
is in state $m$ at time $t$ via the generalized master equation for this probability.  
When $\alpha=1$ the generalized master equation leads at long times 
to an exponential decay of the probability to a steady state, 
exactly as it would for an ordinary Markovian master equation. 
If the WTDs corresponding to the different mechanisms are different, 
the steady state is a nonequilibrium steady state. 
If they are identical, the steady state distribution corresponds 
to equilibrium and satisfies detailed balance. 
On the other hand, if $0<\alpha<1$, the probability $\rho_m(t)$ evolves as a linear 
combination of power law decays $t^{-\alpha}$ towards the zero eigenvalue mode,
and the system never reaches a steady state.

We have presented a formalism to calculate the counting statistics of the energy 
and matter transfer in the CTRW based on a generating function propagation method.
By considering systems exchanging energy and matter with different 
reservoirs, we have shown that $n$th moment of the probability 
distribution to exchange a certain amount of energy or matter
between the system and a given reservoir during a time interval $t$ behaves as
$t^{n \alpha}$ at long times.  This result holds for  $0<\alpha \leq 1$
and reflects the subordination principle~\cite{SokolovKlafter,subordination}.

Using our generating function formalism together with large deviation theory, for WTDs 
with $\alpha=1$ we derived a fluctuation theorem for the trajectory quantity 
$\Delta S_r[m_{\tau}]$ representing the change of entropy along the trajectory due 
to exchange processes with the reservoirs [which can be explicitly related to the 
energy and matter currents between the system and the reservoirs (\ref{FTcurrents})].
If $P(\Delta S_r)$ is the probability to observe a trajectory along which 
occurs a change $\Delta S_r[m_{\tau}]=\Delta S_r$, the fluctuation theorem reads
$P(\Delta S_r)/P(-\Delta S_r)=\exp{\{ \Delta S_r \}}$ and is only valid at long times.
For WTDs with diverging first moments, $0 < \alpha < 1$, this fluctuation theorem does not hold.

The trajectory quantity $R[m_{\tau}]$ is defined as the logarithm of the 
ratio of the probability of a given trajectory to the time reversed trajectory.  
We have shown that for any CTRWs, the ensemble average of $R[m_{\tau}]$ is always 
positive and a fluctuation theorem stating that the probability $P(R)$ to observe 
a trajectory such that $R[m_{\tau}]=R$ is exponentially more likely than to observe 
a trajectory such that $R[m_{\tau}]=-R$, i.e., $P(R)/P(-R)=\exp{\{R\}}$ always holds.  
Separable WTDs are ones for which the state-directional part and the 
temporal part of the WTD factorize [see~(\ref{DTI})].
We have shown that it is only for separable WTDs with $\alpha=1$ that $R[m_{\tau}]$ 
can be interpreted as the change of entropy along the trajectory.
In this case, this fluctuation theorem is related to the previous one because 
$R[m_{\tau}]=\Delta S[m_{\tau}]+\Delta S_r[m_{\tau}]$, where $\Delta S[m_{\tau}]$ 
is the change of system entropy along the trajectory and is bounded, contrary 
to $\Delta S_r[m_{\tau}]$ which typically grows for long trajectories. 
Therefore, for most realizations $R[m_{\tau}] \approx \Delta S_r[m_{\tau}]$ 
at long times, thus providing an interpretation of the fluctuation theorem 
for $\Delta S_r[m_{\tau}]$ in terms of probabilistic asymmetry under time 
reversal of the probability of a trajectory. 
At equilibrium, the symmetry is restored and $R[m_{\tau}]=0$ as expected 
from the principle of microreversibility.
For nonseparable WTDs with $\alpha=1$, $R[m_{\tau}]$ can only be related 
to a change of entropy after a coarse graining of the trajectories in time.
The fluctuation theorem for $\Delta S_r[m_{\tau}]$ still holds, but can only 
be interpreted as a measure of a probabilistic asymmetry in time at the level 
of coarse grained trajectory (and also only for long times), and not at the 
level of the microscopic trajectories as in the case of separable WTDs.  
%%%%%%%%%%%%%%%%%%%%%%%%%%%%%%%%%%%%%%%%%%%%%%%%%%%%%%%%%%%%%%%%%%%%%%
\section*{Acknowledgments}

M. E. would like to thank Franti\v{s}ek \v{S}anda and Shaul Mukamel
for early discussions which motivated this work.
M. E. is supported by the FNRS Belgium (charg\'e de recherches) and 
by the Luxemburgish government (Bourse de formation recherches). 
This research is supported in part by the
National Science Foundation under grant PHY-0354937.

\vskip 10pt
%%%%%%%%%%%%%%%%%%%%%%%%%%%%%%%%%%%%%%%%%%%%%%%%%%%%%%%%%%%%%%%%%%%%%%
\appendix
%%%%%%%%%%%%%%%%%%%%%%%%%%%%%%%%%%%%%%%%%%%%%%%%%%%%%%%%%%%%%%%%%%%%%%
\section{Single level quantum dot}\label{QD}

As an application of the counting statistics results of Sec.~\ref{GFcurrentsec},
we consider a single level quantum 
dot between left and a right leads $\nu=L,R$. 
There are two states in the system, one corresponding to 
the empty level $m=0$ and the other to the filled level $m=1$. 
Since the energy transfer between the system and a
reservoir is directly proportional to 
the particle transfer in this model, we only consider particle transfer.   
We take separable WTDs, specifically
\begin{eqnarray}
&&\psi_{00}(\boldsymbol{\gamma},t) = \psi_{11}(\boldsymbol{\gamma},t) = 0 \\
&&\psi_{10}(\boldsymbol{\gamma},t) = P_{10}(\boldsymbol{\gamma}) \psi_0(t) \nonumber\\
&&\psi_{01}(\boldsymbol{\gamma},t) = P_{01}(\boldsymbol{\gamma}) \psi_1(t) \nonumber \;,
\end{eqnarray}
where 
\begin{eqnarray}
&&P_{10}(\boldsymbol{\gamma}) = 
e^{\gamma^{(L)}} P_{10}^{(L)} + e^{\gamma^{(R)}} P_{10}^{(R)} \nonumber\\
&&P_{01}(\boldsymbol{\gamma}) = 
e^{-\gamma^{(L)}} P_{01}^{(L)}+ e^{-\gamma^{(R)}} P_{01}^{(R)} \;.
\end{eqnarray}
Note that 
\begin{eqnarray}
P_{10}(\boldsymbol{\gamma}=0)=P_{10}^{(L)}+P_{10}^{(R)}=1 \nonumber \\
P_{01}(\boldsymbol{\gamma}=0)=P_{01}^{(L)}+P_{01}^{(R)}=1 \;.
\end{eqnarray}
We define the affinities $A_{x} \equiv \ln P_{01}^{(x)}/P_{10}^{(x)}$, 
where $x=L,R$, and $\Delta A \equiv A_L-A_R$.\\ 

From now on, we focus on the net particle transfer between the 
left lead and the dot so that $\gamma^{(L)}=\gamma$ and $\gamma^{(R)}=0$.
We also note that 
\begin{eqnarray}
&&P_{01}(\gamma) \equiv e^{A_R} P_{10}(\Delta A-\gamma) \nonumber\\
&&P_{10}(\gamma) \equiv e^{-A_R} P_{01}(\Delta A-\gamma) \;. \label{symmetryaffin}
\end{eqnarray}
The generating function (\ref{CTRWGFformalsol}) for this model reads 
\begin{widetext}
\begin{eqnarray}
\tilde{G}(\gamma,s) =
\bigg( \tilde{\phi}_0'(s) + \tilde{\psi}_0'(s) \frac{\tilde{\phi}_1(s) P_{10}(\gamma) 
+ \tilde{\psi}_1(s) \tilde{\phi}_0(s) P_{01}(\gamma) P_{10}(\gamma)}
{1-P_{01}(\gamma) P_{10}(\gamma) \tilde{\psi}_0(s) 
\tilde{\psi}_1(s)} \bigg) \rho_{0}(0) + (1 \leftrightarrow 0) \;.
%\nonumber\\&&
%+\bigg(1+ \frac{\tilde{\phi}_0(s) P_{01}(\gamma) + \tilde{\psi}_0(s) 
%\tilde{\phi}_1(s) P_{01}(\gamma) P_{10}(\gamma)}{1-P_{01}(\gamma) 
%P_{10}(\gamma) \tilde{\psi}_0(s) \tilde{\psi}_1(s)} \bigg) \tilde{\psi}_1'(s)\rho_{1}(0) \;.
\label{Eqtosolve}
\end{eqnarray}
We can verify that the GF at $\gamma=0$ diverge as $s^{-1}$ in the long time limit, 
and, because of (\ref{symmetryaffin}), the same is true at $\gamma=\Delta A$.
The first moment reads
\begin{eqnarray}
\partial_{\gamma} \tilde{G}(\gamma,s)\vert_{\gamma=0} =
\frac{ \tilde{\phi}_0(s) \tilde{\psi}_1(s) \big( P^{(L)}_{10} -P^{(L)}_{01} \big) 
+ \tilde{\phi}_1(s) \big( P^{(L)}_{10}-\tilde{\psi}_0(s) \tilde{\psi}_1(s) P^{(L)}_{01} \big)}
{(\tilde{\psi}_0(s) \tilde{\psi}_1(s)-1)^2} \tilde{\psi}'_0(s) \rho_{0}(0) - (1 \leftrightarrow 0) .
\label{Eqtosolve2}
\end{eqnarray}
\end{widetext}

The WTDs are taken to be of the form
\begin{eqnarray}
\tilde{\psi}_{0}(s)= 1 - B_{0} s^{\alpha} \ \ \;, \ \ 
\tilde{\psi}_{1}(s)= 1 - B_{1} s^{\alpha}
\end{eqnarray}
where $0 \leq \alpha \leq 1$. 
If $\alpha=1$, $B_{0}=t_{0}$ and $B_{1}=t_{1}$.\\

We can now confirm that if $0<\alpha<1$ and a jump occurred at time zero 
($\psi_{m}'(t)=\psi_{m}(t)$), or if $\alpha=1$ and a jump occurred at time 
zero ($\psi_{m}'(t)=\psi_{m}(t)$), or if $\alpha=1$ and we don't know at time zero 
when the last jump occurred ($\tilde{\psi}_{m}'(s)=\tilde{\phi}_{m}(s)/t_{m}$), 
we always get
\begin{eqnarray}
\partial_{\gamma}^n \tilde{G}(\gamma,s)\vert_{\gamma=0} &\stackrel{s\to 0}{=}& 
\frac{n! (P^{(L)}_{10}-P^{(L)}_{01})^n}{(B_0+B_1)^n} s^{-n\alpha-1} ,
\end{eqnarray}
where $n=1,2,3,\ldots$.
Using the Tauberian theorem, this leads to
\begin{eqnarray}
\partial_{\gamma}^n G(\gamma,t)\vert_{\gamma=0} \stackrel{t\to \infty}{=} 
\frac{n! (P^{(L)}_{10}-P^{(L)}_{01})^n}{\Gamma(n \alpha+1) \; (B_0+B_1)^n} \; t^{n \alpha}.
\label{current}
\end{eqnarray}
When $\alpha=1$, (\ref{current}) can be written as
\begin{eqnarray}
&&\hspace{-0.7cm}\partial_{\gamma}^n G(\gamma,t)\vert_{\gamma=0} \stackrel{t\to \infty}{=} 
\bigg( \frac{W_{10}^{L} W_{01}-W_{01}^{L} W_{10}}{W_{10}+W_{01}} \bigg)^n \; t^{n \alpha} 
\label{currentbis} ,
\end{eqnarray}
where $W_{mm'}^{(\nu)}=P_{mm'}^{(\nu)}/t_{m'}$ and $W_{mm'}=W_{mm'}^{(L)}+W_{mm'}^{(R)}$.
This is the same result as obtained in Ref.~\cite{EspositoHarbola07}
using a Markovian master equation.
In can easily be seen that (\ref{currentbis}) vanishes when detailed balance is satisfied.

Finally, we note that the cumulants calculated from these moments all vanish.
This indicates that we should calculate not only the leading time dependence of the moments 
as we do when applying Tauberian theorems, but must retain higher orders to extract information 
about the time dependence of the cumulants.


\begin{thebibliography} {99}


\bibitem{Bustamante02}
J. Liphardt, S. Dumont, S. B. Smith, I. Tinoco (Jr) and C. Bustamante,
%{\it Equilibrium Information from Nonequilibrium Measurements in an Experimental Test of
%Jarzynski's Equality},
Science {\bf 296}, 1832 (2002).

\bibitem{Bustamante04}
E. H. Trepagnier, C. Jarzynski, F. Ritort, G. E. Crooks, C. J. Bustamante, and J. Liphardt,
%{\it Experimental test of Hatano and Sasa’s nonequilibrium steady-state equality},
PNAS {\bf 101}, 15038 (2004).

\bibitem{Bustamante05}
D. Collin, F. Ritort, C. Jarzynski, S.B. Smith, I. Tinoco (Jr), and C. Bustamante,
%{\it Verification of the Crooks fluctuation theorem and recovery of RNA folding free energies}
Nature {\bf 437}, 231 (2005).

\bibitem{Evans05}
G. M. Wang, J. C. Reid, D. M. Carberry, D. R. M. Williams, E. M. Sevick, and D. J. Evans,
%{\it Experimental study of the fluctuation theorem in a nonequilibrium steady state},
Phys. Rev. E {\bf 71}, 046142 (2005).

\bibitem{Seifert05exp}
S. Schuler, T. Speck, C. Tietz, J. Wrachtrup, and U. Seifert,
%{\it Experimental Test of the Fluctuation Theorem for a Driven Two-Level System with Time-Dependent
%Rates},
Phys. Rev. Lett. {\bf 94}, 180602 (2005).

\bibitem{Seifert06exp}
C. Tietz, S. Schuler, T. Speck, U. Seifert, and J. Wrachtrup,
%{\it Measurement of Stochastic Entropy Production},
Phys. Rev. Lett. {\bf 97}, 050602 (2006).

\bibitem{Gaspard07exp}
D. Andrieux, P. Gaspard, S. Ciliberto, N. Garnier, S. Joubaud, and A. Petrosyan,
%{\it Entropy Production and Time Asymmetry in Nonequilibrium Fluctuations}
Phys. Rev. Lett. {\bf 98}, 150601 (2007).

\bibitem{Jarzynski97}
C. Jarzynski, 
%{\it Nonequilibrium Equality for Free Energy Differences},
Phys. Rev. Lett. {\bf 78}, 2690 (1997);
C. Jarzynski, 
%{\it Equilibrium free-energy differences from nonequilibrium measurements: A master-equation approach},
Phys. Rev. E {\bf 56}, 5018 (1997).

\bibitem{VandenBroeck07}
R. Kawai, J. M. R. Parrondo, and C. Van den Broeck, 
%{\it Dissipation: The Phase-Space Perspective},
Phys. Rev. Lett. {\bf 98}, 080602 (2007). 

\bibitem{CleurenVDBroeckKawai}
B. Cleuren, C. Van den Broeck, R. Kawai,
%{\it Fluctuation and dissipation},
C. R. Physique {\bf 8}, 557 (2007).

\bibitem{Kurchan98}
J. Kurchan, 
%{\it Fluctuation theorem for stochastic dynamics},
J. Phys. A {\bf 31}, 3719 (1998).

\bibitem{Lebowitz99}
J. L. Lebowitz and H. Spohn, 
%{\it A Gallavotti–Cohen-Type Symmetry in the Large Deviation Functional for Stochastic Dynamics},
J. Stat. Phys. {\bf 95}, 333 (1999).

\bibitem{Crooks}
G. E. Crooks, 
%{\it Entropy production fluctuation theorem and the nonequilibrium work relation for free energy differences},
Phys. Rev. E {\bf 60}, 2721 (1999); 
%{\it Path-ensemble averages in systems driven far from equilibrium},
Phys. Rev. E {\bf 61}, 2361 (2000).

\bibitem{Seifert05}
U. Seifert, 
%{\it Entropy Production along a Stochastic Trajectory and an Integral Fluctuation Theorem},
Phys. Rev. Lett. {\bf 95}, 040602 (2005).

\bibitem{AndrieuxGaspard07a}
D. Andrieux and P. Gaspard, 
%{\it Fluctuation theorem for currents and Schnakenberg network theory},
J. Stat. Phys. {\bf 127}, 107 (2007).

\bibitem{EspositoHarbola07PRE}
M. Esposito, U. Harbola and S. Mukamel, 
%{\it Entropy fluctuation theorems for driven open systems: application to electron counting statistics},
Phys. Rev. E {\bf 76}, 031132 (2007). 

\bibitem{Evans}
D. J. Evans and D. J. Searles, 
%{\it Steady state fluctuation theorem (nonHamiltonian)
%Equilibrium microstates which generate second law violating steady states},
Phys. Rev. E {\bf 50}, 1645 (1994);
%{\it Steady states, invariant measures, and response theory},
Phys. Rev. E {\bf 52}, 5839 (1995);
%{\it Causality, response theory, and the second law of thermodynamics},
Phys. Rev. E {\bf 53}, 5808 (1996).

\bibitem{Gallavotti}
G. Gallavotti and E. G. D. Cohen, 
%{\it Steady state fluctuation theorem (nonHamiltonian)
%Dynamical Ensembles in Nonequilibrium Statistical Mechanics},
Phys. Rev. Lett. {\bf 74}, 2694 (1995);
G. Gallavotti and E. G. D. Cohen, 
%{\it Dynamical ensembles in stationary states},
J. Stat. Phys. {\bf 80}, 931 (1995).

\bibitem{Kurchan00}
J. Kurchan, 
%{\it A Quantum Fluctuation Theorem},
cond-mat/0007360 (2000).

\bibitem{HTasaki00}
H. Tasaki, 
%{\it Jarzynski relations for quantum systems and some applications},
cond-mat/0009244 (2000).

\bibitem{Mukamel03}
S. Mukamel, 
%{\it Quantum Extension of the Jarzynski Relation: Analogy with Stochastic Dephasing},
Phys. Rev. Lett. {\bf 90}, 170604 (2003).

\bibitem{TalknerHanggi07}
P. Talkner and P. Hanggi, 
%{\it The Tasaki-Crooks quantum fluctuation theorem},
J. Phys. A: Math. Theor. {\bf 40}, F569 (2007).

\bibitem{EspositoHarbola07}
M. Esposito, U. Harbola and S. Mukamel, 
%{\it Fluctuation theorems for counting-statistics in electron transport through quantum junctions},
Phys. Rev. B. {\bf 75}, 155316 (2007).

\bibitem{CleurenVDBroeck}
B. Cleuren and C. Van den Broeck,
%{\it Fluctuation Theorem for Black-body Radiation}, 
Europhys. Lett. {\bf 79}, 30001 (2007).

\bibitem{Kurchan05}
F. Zamponi, F. Bonetto, L. F. Cugliandolo and J. Kurchan,
%{\it A ﬂuctuation theorem for non-equilibrium relaxational systems driven by external forces}
J. Stat. Mech. P09013 (2005). 

\bibitem{Ohta07}
T. Ohkuma and T. Ohta,
%{\it Fluctuation theorems for non-linear generalized Langevin systems}
J. Stat. Mech.  P10010 (2007). 


%\bibitem{Gustavsson06}
%S. Gustavsson, R. Leturcq, B. Simovic, R. Schleser, T. Ihn, 
%P. Studerus, K. Ensslin, D. C. Driscoll, and A. C. Gossard, 
%{\it Counting Statistics of Single Electron Transport in a Quantum Dot},
%Phys. Rev. Lett. {\bf 96}, 076605 (2006).

%\bibitem{Hirayama06}
%T. Fujisawa, T. Hayashi, R. Tomita, and Y. Hirayama, 
%{\it Bidirectional Counting of Single Electrons}
%Science {\bf 312}, 1634 (2006).

\bibitem{Montroll65}
E. W. Montroll and G. H. Weiss,
%{\it Random Walks on Lattices. II}
J. Math. Phys. {\bf 6}, 167 (1965).

\bibitem{Lax73}
H. Scher and M. Lax, Phys. Rev. B {\bf 7}, 4491 (1973); {\bf 7}, 4502 (1973).

\bibitem{Kehr87}
J. W. Haus and K. W. Kehr, Phys. Rep. {\bf 150}, 263 (1987). 

\bibitem{Georges90}
J.-P. Bouchard and A. Georges, Phys. Rep. 195, 127 (1990). 

\bibitem{Lindenberg71}
D. Bedeaux, K. Lakatos-Lindenberg and K. Shuler, J. Math. Phys. {\bf 12}, 2116 (1971).

\bibitem{Hughes}
B. D. Hughes,
{\it Random Walks and Random Environments: Volume 1: Random Walks}, 
(Oxford University Press, USA, 1995);
{\it Random Walks and Random Environments: Volume 2: Random Environments} 
(Oxford University Press, USA, 1996).

\bibitem{Schnakenberg}
J. Schnakenberg,
%{\it Network theory of microscopic and microscopic behaviour of master equation systems},
Rev. Mod. Phys. {\bf 48}, 571 (1976).

\bibitem{Barkai03}
E. Barkai and Y. C. Cheng, 
%{\it Aging Continuous Time Random Walks},
J. Chem. Phys. {\bf 118}, 6167 (2003). 

\bibitem{Allegrini03}
P. Allegrini, G. Aquino, P. Grigolini, L. Palatella, and A. Rosa,
%{\it Generalized master equation via aging continuous-time random walks}
Phys. Rev. E {\bf 68}, 056123 (2003).

\bibitem{Sokolov}
W. Ebeling and I. M. Sokolov,
{\it Statistical thermodynamics and stochastic theory of nonequilibrium systems},
(World Scientific, 2005).

\bibitem{Barkai06}
G. Bel and E. Barkai,  
%{\it Random walk to a nonergodic equilibrium concept},
Phys. Rev. E {\bf 73}, 016125 (2006).

\bibitem{Barkai07}
A. Rebenshtok and E. Barkai,
%{\it Distribution of Time-Averaged Observables for Weak Ergodicity Breaking},
Phys. Rev. Lett. {\bf 99}, 210601 (2007).

\bibitem{Silbey02}
Y. Jung, E. Barkai and R. J. Silbey
%{\it Lineshape theory and photon counting statistics for blinking quantum dots: a L\'evy walk process}
Chem. Phys. {\bf 284}, 181 (2002).

\bibitem{Bouchaud03}
X. Brokmann, J.-P. Hermier, G. Messin, P. Desbiolles, J.-P. Bouchaud, and M. Dahan,
%{\it Statistical Aging and Nonergodicity in the Fluorescence of Single Nanocrystals}
Phys. Rev. Lett. {\bf 90}, 120601 (2003).

\bibitem{Yang03}
H. Yang, G. Luo, P. Karnchanaphanurach, T. M. Louie, I. Rech, S. Cova, L. Xun, X. S. Xie,
%{\it Protein Conformational Dynamics Probed by Single-Molecule Electron Transfer}
Science {\bf 302}, 262 (2003).

\bibitem{KlafterExp05}
O. Flomenbom, K. Velonia, D. Loos, S. Masuo, M. Cotlet, Y. Engelborghs, J. Hofkens, A.E.
Rowan, R.J.M. Nolte, M. Van der Auweraer, F.C. De Schryver, and J. Klafter, 
%{\it Stretched exponential decay and correlations in the catalytic activity 
%of fluctuating single lipase molecules}
PNAS {\bf 102}, 2368 (2005).

\bibitem{Shlesinger}
M. F. Shlesinger,
%{\it Asymptotic Solutions of Continuous-Time Random Walks},
J. Stat. Phys., {\bf 10}, 5 (1974).

\bibitem{Metzler}
R. Metzler and J. Klafter,
%{\it The random walk's guide to anomalous diffusion: a fractional dynamics approach}
Phys. Rep. {\bf 339}, 1 (2000).

\bibitem{Feller}
W. Feller, {\it An Introduction to Probability Theory and its Applications}, 
2nd ed. (Wiley, New York, 1971), Vol II.


\bibitem{Korevaar}
J. Korevaar, {\it Tauberian Theory: A century of developments} (Springer, Berlin, 2004).
%%%%%%

\bibitem{WangQian}
H. Wang and H. Qian,
%{\it On detailed balance and reversibility of semi-Markov processes and single-molecule enzyme kinetics},
J. Math. Phys. {\bf 48}, 013303 (2007). 

\bibitem{QianWang}
H. Qian and H. Wang,
%{\it Continuous time random walks in closed and open single-molecule systems with microscopic reversibility},
Europhys. Lett. {\bf 76}, 15 (2006). 

\bibitem{SokolovKlafter} I. M. Sokolov and J. Klafter, Chaos {\bf 15}, 026103 (2005).

\bibitem{subordination} M. M. Meerschaert, D. A. Benson, H.-P. Scheffler and B. Baeumer, Phys. Rev.
E {\bf 65}, 041103 (2002); A. A. Satinislavsky, Physica A {\bf 318}, 469 (2003); R. Gorenflo,
F. Mainardi and A. Vivoli, Chaos, Sol. Frac. {\bf 34}, 89 (2007).

\end{thebibliography}
\end{document}